\documentclass[aps,showpacs,preprintnumbers,amsmath,amssymb,nofootinbib]{revtex4}
\usepackage{amssymb}
\usepackage{amsfonts}
\usepackage{amsmath}
\usepackage{graphicx}
\usepackage{dcolumn}
\usepackage{bm}
\usepackage[latin1]{inputenc}
\usepackage{graphicx}
\def\be{\begin{equation}}
 \def\ee{\end{equation}}
 \def\bea{\begin{eqnarray}}
 \def\eea{\end{eqnarray}}

\def\PR#1{{Phys.\ Rev.\ D \bf #1}}
\def\PRL#1{{Phys.\ Rev.\ Lett.\ \bf #1}}

\begin{document}

\title{ Extremal Hairy Black Holes}
\author{P. A. Gonz\'{a}lez}
\email{\,\,\,\,pablo.gonzalez@udp.cl} \affiliation{Facultad de
Ingenier\'{i}a, Universidad Diego Portales, Avenida Ej\'{e}rcito
Libertador 441, Casilla 298-V, Santiago, Chile.}
\author{Eleftherios Papantonopoulos}
\email{\,\,\,\,lpapa@central.ntua.gr} \affiliation{Department of
Physics, National Technical University of Athens, Zografou Campus
GR 157 73, Athens, Greece.}
\author{Joel Saavedra}
\email{\,\,\,\,joel.saavedra@ucv.cl} \affiliation{Instituto de
F\'{i}sica, Pontificia Universidad Cat\'olica de Valpara\'{i}so,
Casilla 4950, Valpara\'{i}so, Chile.}
\author{Yerko V\'{a}squez}
\email{\,\,\,\,yvasquez@userena.cl}
\affiliation{Departamento de F\'{\i}sica, Facultad de Ciencias, Universidad de La Serena,\\
Avenida Cisternas 1200, La Serena, Chile.}
\date{\today}

\begin{abstract}
We consider a gravitating system consisting of a scalar field
minimally coupled to gravity with a self-interacting potential and
an U(1) electromagnetic field. Solving the coupled
Einstein-Maxwell-scalar system we find exact hairy charged black
hole solutions with the scalar field regular everywhere. We go to
the zero temperature limit and we study the effect of the scalar
field on the near horizon geometry of an extremal black hole. We
find that except a critical value of the charge of the black hole
there is also a critical value of the charge of the scalar field
beyond of which the extremal black hole is destabilized. We study
the thermodynamics of these solutions and we find that if the
space is flat then the Reissner-Nordstr\"om
 black hole is thermodynamically preferred, while if the space
is AdS the hairy charged black hole is thermodynamically preferred
at low temperature.
\end{abstract}

\maketitle
\section{introduction}
\label{secs.1}

The gauge/gravity duality is a principle that it is well founded
in string theory and connects a strongly coupled d-dimensional
conformal field theory with a (d+1)-dimensional gravity theory
that is weakly coupled \cite{Maldacena:1997re}. This principle has
been applied to many field theories having gravity duals but its
most noticeable application is in condensed matter physics.
Recently there is a lot of activity in understating the properties
of quantum liquids \cite{Liu:2009dm,Faulkner:2009wj}. Quantum
liquids arise if we put a many-body system of a finite $U(1)$
charge density at zero temperature. Understanding the ground
states of these finite density systems at strong coupling will
give us information about the nature of these quantum liquids and
will lead us to find applications in condensed matter systems.

According to gauge/gravity duality the gravity part of these
systems is  described by an extremal charged black hole in anti-de
Sitter spacetime \cite{Chamblin:1999tk}.  The metric of the
extremal black hole has  interesting properties. In the near
horizon limit when the temperature goes to zero the horizon
geometry is given by AdS$_2 \times R^{2}$ \cite{Kunduri:2013gce}.
This happens because  the charge of the black hole introduces
another scale. The appearance of a new horizon geometry suggests
that the boundary system could develop an enhanced symmetry. The
other property is that the black hole has a finite horizon area at
zero temperature. So we can assign a non-zero entropy at zero
temperature.

The properties of the near horizon geometry were also used to
explore the zero temperature limit of holographic fluids and
superconductors \cite{Horowitz:2009ij}. It was shown in
\cite{Alsup:2011qn} that there is a critical temperature $T_c$
where a charged scalar field condenses, and as $T_c \rightarrow 0$
there is a critical value $q_c^2$ of the charge of the scalar
field attained in the zero temperature limit   and this limit is
determined solely by the AdS$_2$ geometry of the horizon. In
connection to holographic superconductors the near horizon
geometry of an extremal black hole in the presence of charged
scalar fields was studied in \cite{FernandezGracia:2009em}.
Conditions for the existence of scalar hair of neutral and charged
scalar fields were derived. Also exact hairy black hole solutions
were found without an electromagnetic field  \cite{Zeng:2009fp}.

These developments put forward the necessity  of a better
understanding of the behaviour of matter fields near  the  horizon
of a charged black hole at the zero temperature limit. This can be
achieved if one has a fully back-reacted solution of the
Einstein-Maxwell-scalar system. The main difficulty for such a
construction is to evade the no-hair theorems and have  a healthy
behaviour of the scalar field: regular on the horizon and  fall
off sufficiently fast at large distances. The aim of this work is
to study the properties and the behaviour of the fields in the
near horizon geometry of a charged black hole as the temperature
goes to zero. To achieve  this we will use a profile for the
scalar field and we will probe the near horizon geometry solving
exactly the Einstein-Maxwell-scalar coupled differential
equations. Exact solutions of this system without the
electromagnetic field were found in \cite{Gonzalez:2013aca}.

Hairy black holes are interesting solutions of Einstein's Theory
of Gravity and
 have been extensively studied over the years mainly in connection
 with the no-hair theorems. Then hairy black hole solutions
were found  in  asymptotically flat spacetimes \cite{BBMB} but
 it was realized that these solutions  were not physically acceptable as the scalar field
was divergent on the horizon and stability analysis showed that
they were unstable \cite{bronnikov}. To remedy this a
regularization procedure has to be used to make the scalar field
finite on the horizon.

The easiest way to make the scalar field regular on the horizon is
to introduce a scale in the gravity sector of the theory through a
cosmological constant. Then various hairy black hole solutions
were found \cite{Zloshchastiev:2004ny}-\cite{Martinez:2004nb}. A
characteristic of these solutions is  that the parameters
connected with the scalar fields are connected in some way with
the physical parameters of the hairy solution. This implies that
it is not possible to continuously connect the hairy configuration
with mass $M$ and a configuration with the same mass and no scalar
field.

Hairy solutions were also found of a scalar field coupled to a
charged black holes. In \cite{Martinez:2005di} a topological black
hole dressed with a conformally coupled scalar field and
  electric charge was studied. Phase transitions of hairy topological black holes were
  studied in \cite{Koutsoumbas:2006xj,Martinez:2010ti}. An electrically charged black hole solution with a scalar
  field minimally coupled to gravity and electromagnetism
 was presented in \cite{Martinez:2006an}. It was found that regardless the value of the electric charge,
the black hole is massless and has a fixed temperature. The
thermodynamics of the solution was also studied. Further hairy
solutions  were reported in
\cite{Kolyvaris:2009pc}-\cite{Charmousis:2014zaa} with various
properties. More recently new hairy black hole solutions, boson
stars and numerical rotating hairy black hole solutions were
discussed
\cite{Dias:2011at,Stotyn:2011ns,Dias:2011tj,Kleihaus:2013tba,Herdeiro:2014goa,Herdeiro:2014ima}.
Also the thermodynamics of hairy black holes was studied in
\cite{Lu:2014maa}.

In spite of this progress little are known on the behaviour of
hairy black holes as the temperature goes to zero. To probe the
near horizon limit of a charged black hole we introduce a profile
of the scalar field that it falls off sufficiently fast outside
the horizon. Then by solving the coupled Einstein-Maxwell-scalar
system we find exact hairy charged black hole solutions with the
scalar field regular everywhere. Then we go to the zero
temperature limit and we study the effect of the scalar field on
the near horizon geometry of an extremal black hole. We find that
except a critical value of the charge of the black hole there is
also a critical value of the charge of the scalar field away of
which the extremal black hole is destabilized. We also study the
thermodynamics of these solutions and we find that if the space is
flat then the Reissner-Nordstr\"om (RN) black
hole is thermodynamically preferred, while if the space is AdS the
hairy charged black hole is thermodynamically preferred at low
temperature.

 The work is organized as follows. In Section (\ref{secs.2}) we
 present the general formalism and we derive  the field equations.
 In  Section (\ref{secs.3}) we find exact hairy black hole
 solutions and we study their properties. In Section (\ref{secs.4})
 we study the effect of the scalar field on the near horizon
 geometry. In Section (\ref{secs.5}) we study the thermodynamics of
 our solutions while in (\ref{secs.6}) are our conclusions.

\section{General Formalism}
\label{secs.2}

In this section we will review the general formalism discussed in
\cite{Gonzalez:2013aca} of a scalar field minimally coupled to
curvature having a self-interacting potential $ V(\phi)$, in the
presence of an electromagnetic field. The Einstein-Hilbert action
with  a negative cosmological constant $\Lambda=-6l^{-2}/\kappa$,
where $l$ is the length of the AdS which has been incorporated in
the potential as $ \Lambda=V(0)$ ($V(0)<0$) is
 \begin{eqnarray} \label{action}
 S=\int d^{4}x\sqrt{-g}\left(\frac{1}{2 \kappa }R-\frac{1}{4}F_{\mu \nu}F^{\mu \nu}
 -\frac{1}{2}g^{\mu\nu}\nabla_{\mu}\phi\nabla_{\nu}\phi-V(\phi)\right)~,
 \end{eqnarray}
where $\kappa=8 \pi G_N$, with $G_N$ the Newton constant. The
resulting Einstein equations from the above action are
 \begin{eqnarray}
 R_{\mu\nu}-\frac{1}{2}g_{\mu\nu}R=\kappa (T^{(\phi)}_{\mu\nu}+T^{(F)}_{\mu\nu})~,\label{field1}
 \end{eqnarray}
the energy momentum tensors $T^{(\phi)}_{\mu\nu}$ and $T^{(F)}_{\mu\nu}$ for the
scalar and electromagnetic fields are
 \begin{eqnarray}
 \nonumber T^{(\phi)}_{\mu\nu}&=&\nabla_{\mu}\phi\nabla_{\nu}\phi-
 g_{\mu\nu}[\frac{1}{2}g^{\rho\sigma}\nabla_{\rho}\phi\nabla_{\sigma}\phi+V(\phi)]~,\\
  T^{(F)}_{\mu\nu}&=&F_{\mu}^{\alpha}F_{\nu \alpha} -\frac{1}{4}g_{\mu \nu}F_{\alpha\beta}F^{\alpha\beta}  \label{energymomentum}~.
 \end{eqnarray}
 If we use Eqs. (\ref{field1}) and (\ref{energymomentum}) we obtain the equivalent equation
 \begin{eqnarray}
 R_{\mu\nu}-\kappa\left(\partial_\mu\phi \partial_\nu\phi+g_{\mu\nu}V(\phi)\right)=\kappa (F_{\mu}^{\alpha}F_{\nu \alpha} -\frac{1}{4}g_{\mu \nu}F_{\alpha\beta}F^{\alpha\beta})~. \label{einstein1}
 \end{eqnarray}
We consider the following metric ansatz
 \begin{eqnarray}
 ds^{2}=-f(r)dt^{2}+f^{-1}(r)dr^{2}+a^{2}(r)d \sigma^2 ~,\label{metricBH}
 \end{eqnarray}
where $d \sigma ^2$ is the metric of the spatial 2-section, which
can have positive, negative or zero  curvature, and
$A_{\mu}=(A_t(r),0,0,0)$ the scalar potential of the
electromagnetic field. In the case of the metric of Eq.
(\ref{metricBH}), if we use Eq. (\ref{einstein1}) we find the
following three independent differential equations
 \begin{eqnarray}
 f''(r)+2\frac{a'(r)}{a(r)}f'(r)+2V(\phi)=A^{\prime}_t(r)^2~,\label{first}
 \end{eqnarray}
\begin{eqnarray}
\frac{a'(r)}{a(r)}f'(r)+\left(\left(\frac{a'(r)}{a(r)}\right)^{2}+\frac{a''(r)}{a(r)}\right)f(r)-\frac{k}{a(r)^{2}}+V(\phi)=-\frac{1}{2}A^{\prime}_t(r)^2~,\label{second}
\end{eqnarray}
\begin{eqnarray}
f''(r)+2\frac{a'(r)}{a(r)}f'(r)+\left(4\frac{a''(r)}{a(r)}+2(\phi'(r))^{2}\right)f(r)+2V(\phi)=A^{\prime}_t(r)^2~,\label{third}
\end{eqnarray}
where $k$ is the curvature of the spatial 2-section.  All the
quantities, in the above equations, have been rendered
dimensionless via the redefinitions $\sqrt{\kappa} \phi\rightarrow
\phi$ and $\kappa  V \rightarrow V$. Now, if we eliminate the potential $V(\phi)$ from the above equations
we obtain
\begin{eqnarray}
a''(r)+\frac{1}{2}(\phi'(r))^2 a(r)=0\label{adiff}~,
\end{eqnarray}
\begin{eqnarray}
f''(r)-2\left(
\left(\frac{a'(r)}{a(r)}\right)^{2}+\frac{a''(r)}{a(r)}\right)f(r)+\frac{2k}{a(r)^2}=2A^{\prime}_t(r)^2~,\label{fdiff}
\end{eqnarray}
where the potential can be determined from Eq. (\ref{first}) if
the functions $a(r)$ and $f(r)$ are known. To find exact hairy
black hole solutions  the differential equations
(\ref{first})-(\ref{third}) have to be supplemented with the
Klein-Gordon equation of the scalar field and the Maxwell
equations which in general coordinates read
\begin{eqnarray}\label{klg}
\nonumber  \Box \phi &=&\frac{d V}{d \phi}~,\\
\nabla_{\nu}F^{\mu\nu}&=&0~.
\end{eqnarray}

\section{A four-dimensional charged black holes with scalar hair}
\label{secs.3}

Following the general formalism developed in Section II for a scalar
field coupled minimally to gravity, we consider a particular
profile of the scalar field. Consider the following ansatz for the
scalar field
\begin{equation}
\phi \left( r\right) =\frac{1}{\sqrt{2}}\ln \left( 1+\frac{\nu }{r}\right) ~,
\label{field}
\end{equation}
where $\nu $ is a parameter controlling the behaviour of the
scalar field and it has the dimension of length.   Then from
equation (\ref{adiff}) and (\ref{klg}) we can determine the
functions
\begin{eqnarray}\label{ametric}
\nonumber a\left( r\right)& =&\sqrt{r\left( r+\nu \right) }~,\\
A_t(r)&=&\frac{q}{\nu}ln\left(\frac{r}{r+\nu}\right)~,
\end{eqnarray}
analytically. We can also
determine the metric function $f(r)$ analytically using equation
(\ref{fdiff}). We find
\begin{eqnarray}
 f\left( r\right) &= &-2\frac{q^2}{\nu^2}+C_1r(r+\nu)-
\frac{C_2(2r+\nu)}{\nu^2}+2\frac{k r(2r+\nu)} {\nu^2} \nonumber
\\
&& -2\left(\frac{ q^2(2r+\nu)+r(r+\nu)(C_2+k\nu)
}{\nu^3}+\frac{q^2r(r+\nu)ln\frac{r}{r+\nu}}{\nu^4}\right)
ln\frac{r}{r+\nu}~, \label{f(r)}
\end{eqnarray}
where $k=-1,0,1$ and $C_1$, $C_2$ are  integration constants being proportional to
the cosmological constant and to the mass respectively,
and the potential is given by
\begin{eqnarray}
\nonumber V\left( \phi \right) &=&  \frac{1}{2\nu^4}e^{-2\sqrt{2}\phi}\left[\left(e^{\sqrt{2}\phi}-1\right)^2\left(1+10e^{\sqrt{2}\phi} +e^{2\sqrt{2}\phi}\right)q^2 \right. \nonumber
\\
&&
+e^{\sqrt{2}\phi}\nu \left(-6C_2-10k\nu-C_1\nu^3-4e^{\sqrt{2}\phi}\nu\left(4k+C_1\nu^2\right)+e^{2\sqrt{2}\phi}\left(6C_2+2k\nu-C_1\nu^3\right)\right)\nonumber
\\
&&
+2\sqrt{2}\phi\left(\left(1+4e^{\sqrt{2}\phi}+e^{2\sqrt{2}\phi}\right)q^2ln\frac{\nu}{e^{\sqrt{2}\phi-1}} \right.\nonumber
\\
&&
\left. \left. +2e^{\sqrt{2}\phi} \left( \left(2+cosh\sqrt{2}\phi \right)\left( \nu C_2+k\nu^2-q^2ln\frac{e^{\sqrt{2}\phi}\nu}{e^{\sqrt{2}\phi-1}}\right)+6q^2sinh\sqrt{2}\phi\right) \right)\right],
\label{m2potn}
\end{eqnarray}
where $ V\left( 0\right) =\Lambda_{eff}$  as expected and also
\begin{equation} C_1+\frac{4k}{\nu^2}=-\frac{\Lambda_{eff}
}{3}=\frac{1}{l^{2}}~. \label{relation}\end{equation} We see from
the above relation that the parameter $\nu$ of the scalar field
introduces a length scale connected with the presence of the
scalar field in the theory.
Besides, we know that%
\begin{equation}
V^{\prime \prime }\left( \phi =0\right) =m^{2}~,
\end{equation}
where $m$ is the scalar field mass. Therefore, we obtain that the
scalar field mass is given by
\begin{equation}
m^{2}=\frac{2}{3}\Lambda_{eff} =-2l^{-2}~,
\end{equation}
which satisfies the Breitenlohner-Friedman bound that ensures the
perturbative stability of the AdS spacetime
\cite{Breitenlohner:1982jf}.

One may wander if in the limit of $\Lambda_{eff} \rightarrow 0$
and $\nu \rightarrow 0$ we recover the Reissner-Nordstrom (RN)
black hole. Indeed from  (\ref{f(r)}) if we fix the constant $C_1$
to $C_1=-\frac{4k}{\nu^2}$  the function $f(r)$ can be written as
\begin{eqnarray}
f(r)&=& -\frac{2\left(q^2+5\nu+r(C_2+k\nu)\right)}{\nu^2} \nonumber
\\
&&-\frac{2\left(\nu\left(q^2(2r+\nu)+r(r+\nu)(C_2+k\nu)\right)+q^2r(r+\nu)ln\frac{r}{r+\nu}\right)ln\frac{r}{r+\nu}}{\nu^4}~,
\label{RNL1}
\end{eqnarray}
and in the limit  $\nu \rightarrow 0$ we recover the RN black hole
\begin{equation}\label{RNL2}
f(r)=k+\frac{q^2}{2r^2}-\frac{C_2}{3r}~.
\end{equation}

It is interesting to investigate the case with $\nu \neq 0$. To
have a better understanding of the resulting geometry we make a
change of coordinates $\rho =\sqrt{r\left( r+\nu \right) }$. Then
 the metric (\ref{f(r)}) can be written as
\begin{equation}
ds^{2}=-\chi \left( \rho \right) dt^{2}+\frac{4\rho ^{2}/\nu
^{2}}{4\rho ^{2}/\nu ^{2}+1}\frac{1}{\chi \left( \rho \right)
}d\rho ^{2}+\rho ^{2}d\sigma ^{2}~, \label{metricnew}
\end{equation}
where
\begin{eqnarray}
\nonumber \chi \left( \rho
\right)&=&k-\frac{2q^2}{\nu^2}+\rho^2\left(C_1+\frac{4k}{\nu^2}\right)-\left(k+\frac{C_2}{\nu
}\right)\left( \sqrt{\frac{4\rho ^{2}}{\nu ^{2}}+1}-2\frac{\rho
^{2}}{\nu ^{2}}\ln \left( \frac{1+\sqrt{
\frac{4\rho ^{2}}{\nu ^{2}}+1}}{-1+\sqrt{\frac{4\rho ^{2}}{\nu ^{2}}+1}}\right) \right)\\
&&+ \frac{2q^2}{\nu^2}ln \left( \frac{1+\sqrt{\frac{4\rho ^2}{\nu
^2}+1}}{-1+\sqrt{\frac{4\rho ^2}{\nu ^2}+1}}\right)\left(
\sqrt{\frac{4\rho ^{2}}{\nu ^{2}}+1}-\frac{\rho ^{2}}{\nu ^{2}}\ln
\left( \frac{1+\sqrt{ \frac{4\rho ^{2}}{\nu
^{2}}+1}}{-1+\sqrt{\frac{4\rho ^{2}}{\nu ^{2}}+1}}\right)
\right)~.
\end{eqnarray}
The scalar field in the new coordinates reads
\begin{equation}
\phi \left( \rho \right) =\frac{1}{\sqrt{2}}\ln \left( \frac{1+\sqrt{\frac{%
4\rho ^{2}}{\nu ^{2}}+1}}{-1+\sqrt{\frac{4\rho ^{2}}{\nu
^{2}}+1}}\right)~.\label{scafiel}
\end{equation}
Then it is clear that the cosmological constant is modified having
contributions from the length scale introduced by the scalar
field.

At  large distances the scalar field decouples and the metric goes
to
\begin{equation}
\chi \left( \rho \right)
=\left(C_1+\frac{4k}{\nu^2}\right)\rho^2+k-
\frac{C_2+k\nu}{3\rho}+\frac{q^2}{2\rho^2}+\mathcal O\left(
\frac{1} {\rho ^{3}}\right)~ .
\end{equation}
From the above relation we can see that the asymptotic behaviour
can be RN anti-de Sitter, RN de Sitter or RN metric by depending
of the sign of the term proportional to the effective cosmological
constant $-\frac{\Lambda_{eff}}{3}=C_1+\frac{4k}{\nu^2}$, as expected.

\begin{figure}[h]
\begin{center}
\includegraphics[scale=.7, natwidth=0.4, natheight=0.5]{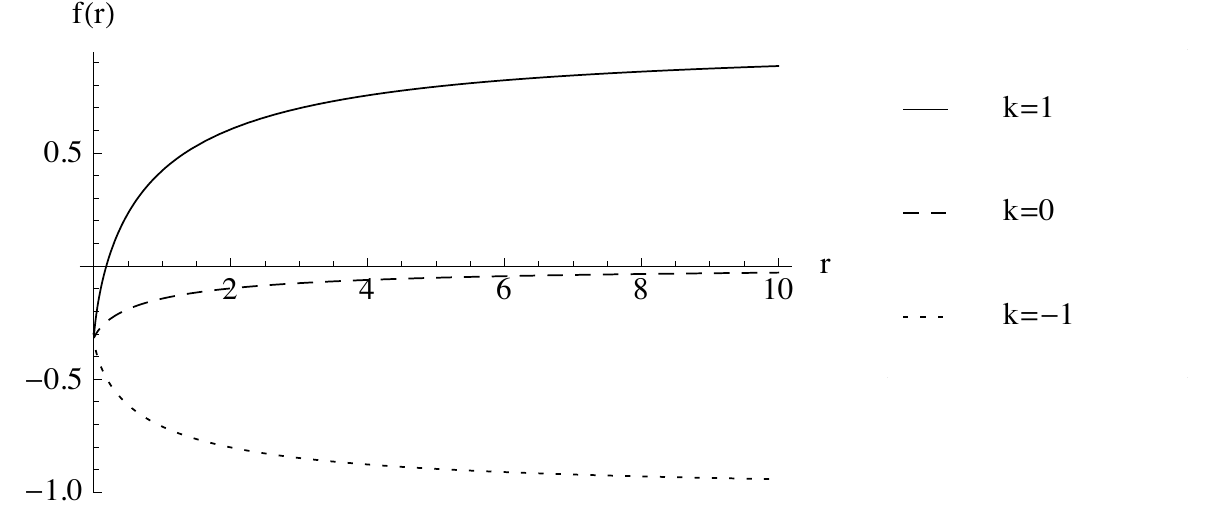}
\includegraphics[scale=.7, natwidth=0.4, natheight=0.5]{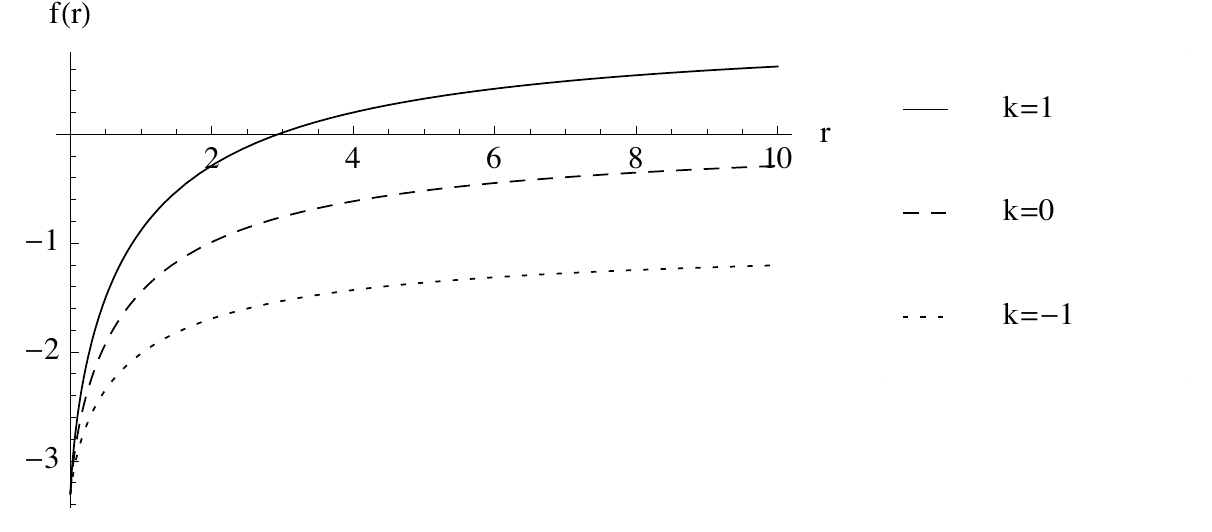}
\includegraphics[scale=.7, natwidth=0.4, natheight=0.5]{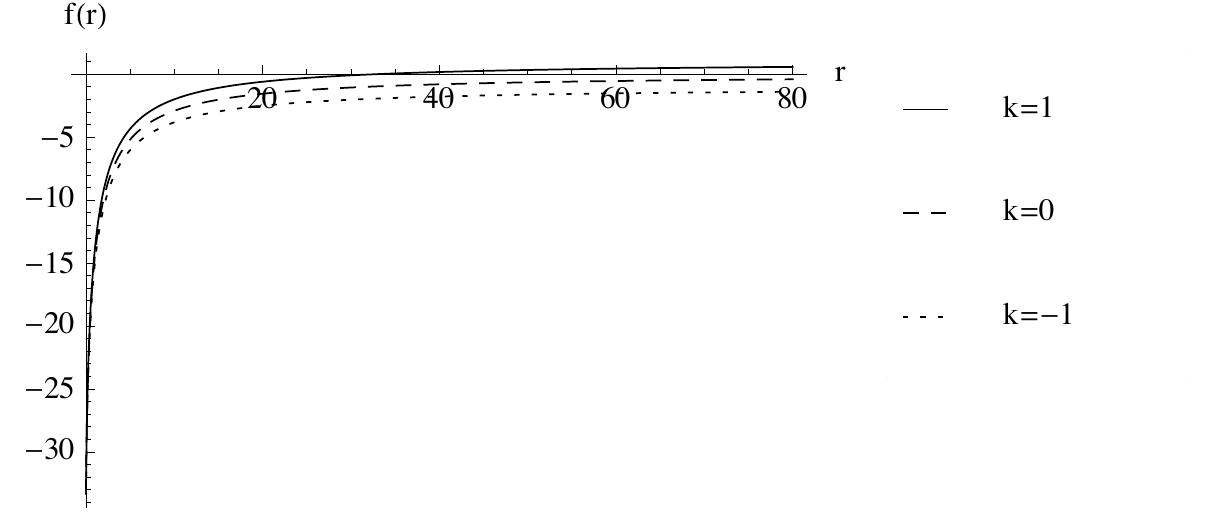}
\end{center}
\caption{The behaviour of $f(r)$, for $\protect%
\nu =3$, $q=0.1$, $C_2=1$ left figure, $C_2=10$ right figure, and $C_2=100$ bottom figure.}
\label{plots0}
\end{figure}

\begin{figure}[h]
\begin{center}
\includegraphics[scale=.7, natwidth=0.4, natheight=0.5]{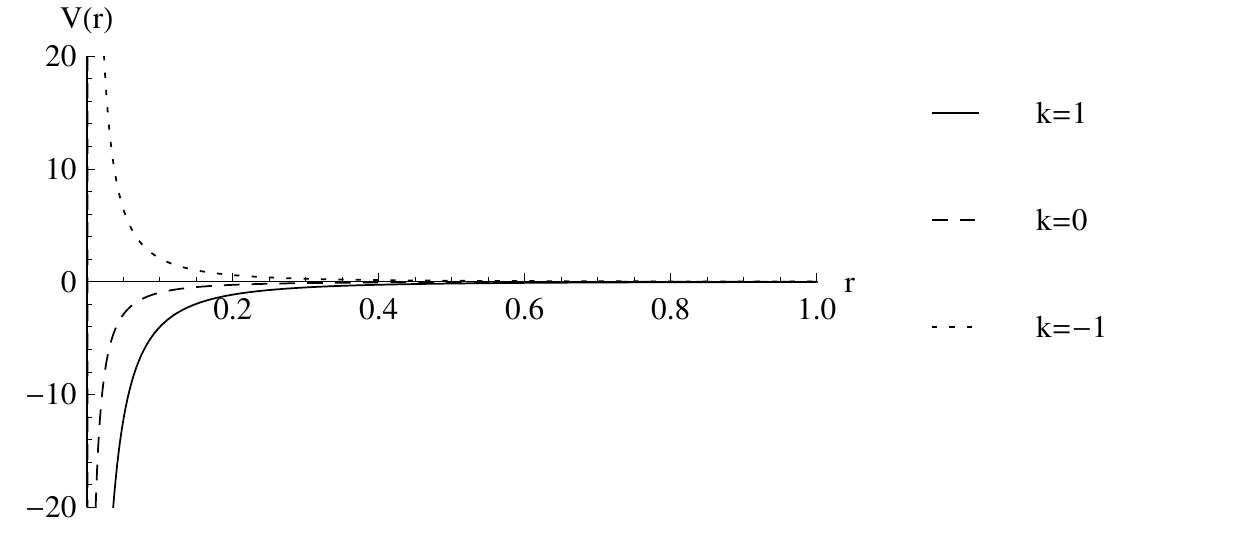}
\includegraphics[scale=.7, natwidth=0.4, natheight=0.5]{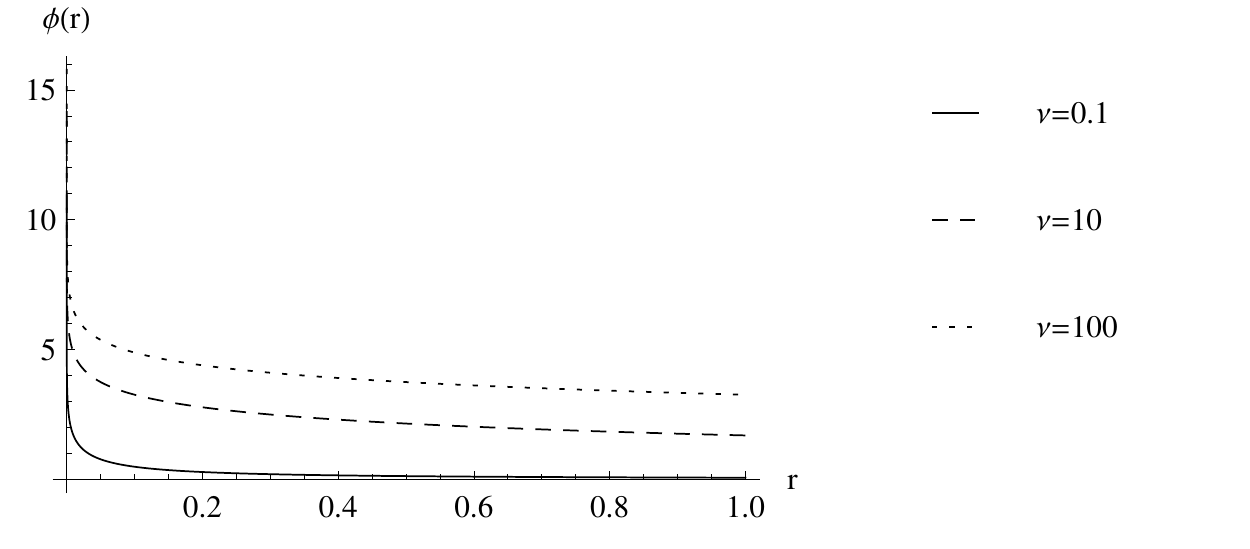}
\end{center}
\caption{Left figure corresponds to the behaviour of $V(r)$, for $\protect%
\nu =3$, $C_2=1$, $q=0.1$, and $k=1$, $0$, $-1$. Right figure
corresponds to the behaviour of $\phi(r)$ for  $C_2=1$, $q=0.1$,
and $\nu =0.1, 3, 10$, and $100$.} \label{plots11}
\end{figure}

\begin{figure}[h]
\begin{center}
\includegraphics[scale=.7, natwidth=0.4, natheight=0.5]{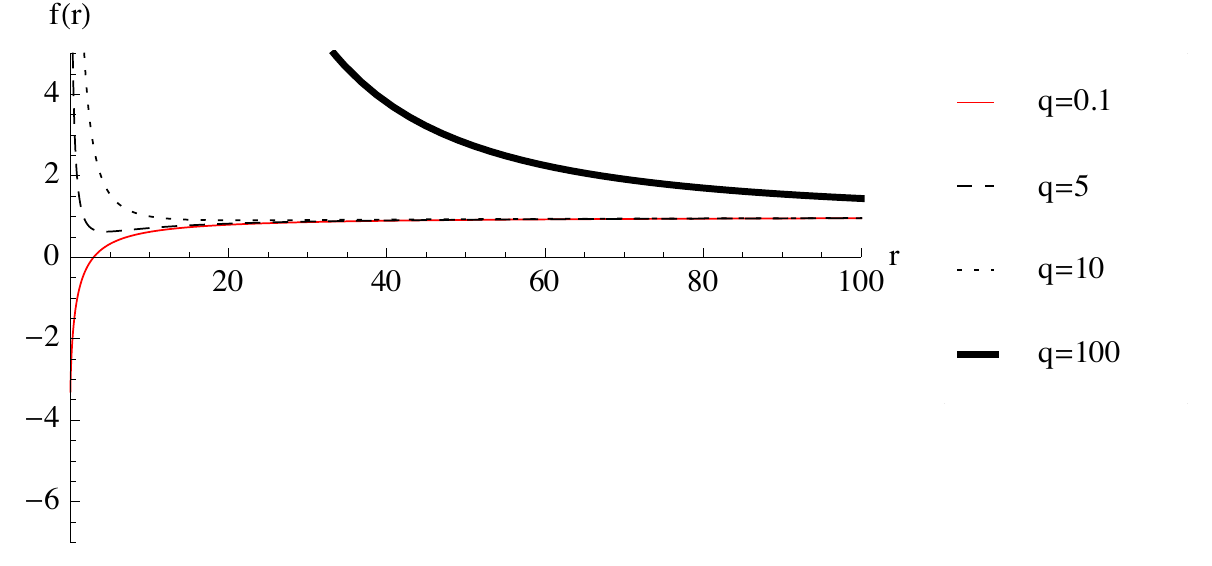}
\includegraphics[scale=.7, natwidth=0.4, natheight=0.5]{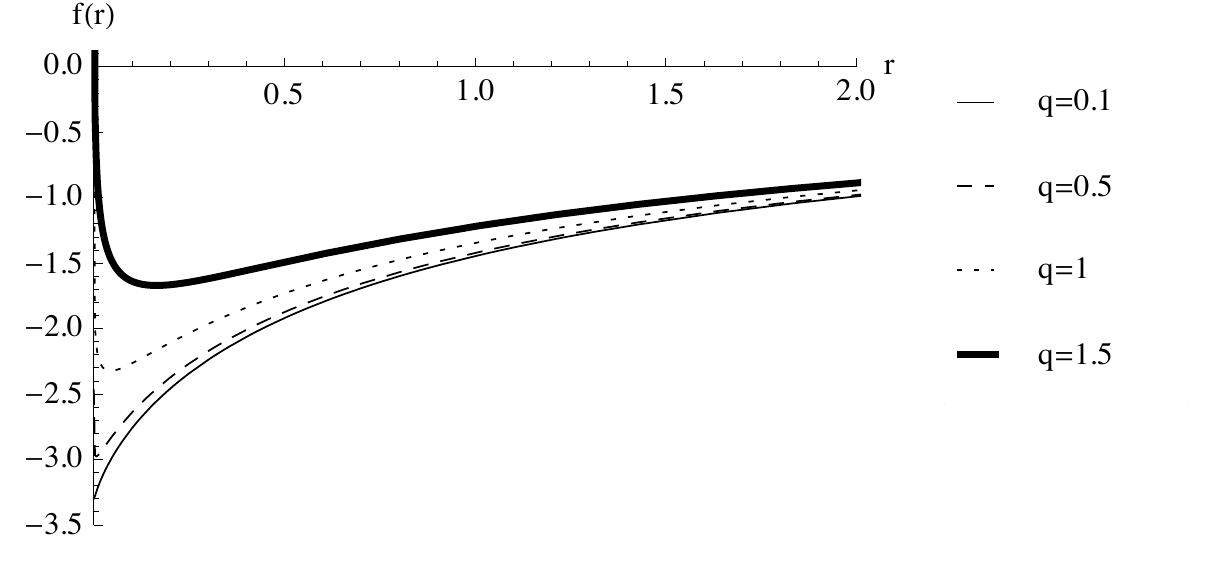}
\includegraphics[scale=.7, natwidth=0.4, natheight=0.5]{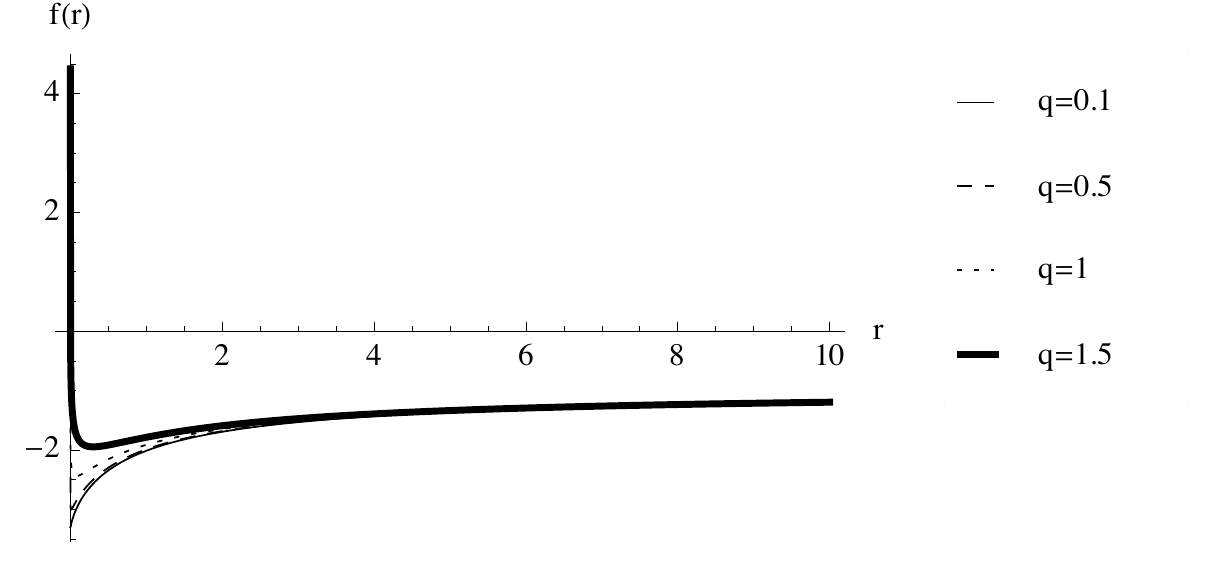}
\end{center}
\caption{The behaviour of $f(r)$, for $\protect%
\nu =3$, $C_2=10$ and $k=1$ left figure, $k=0$ right figure, and $k=-1$ bottom figure for different values of $q$.} \label{plotsR1}
\end{figure}

\begin{figure}[h]
\begin{center}
\includegraphics[scale=.7, natwidth=0.4, natheight=0.5]{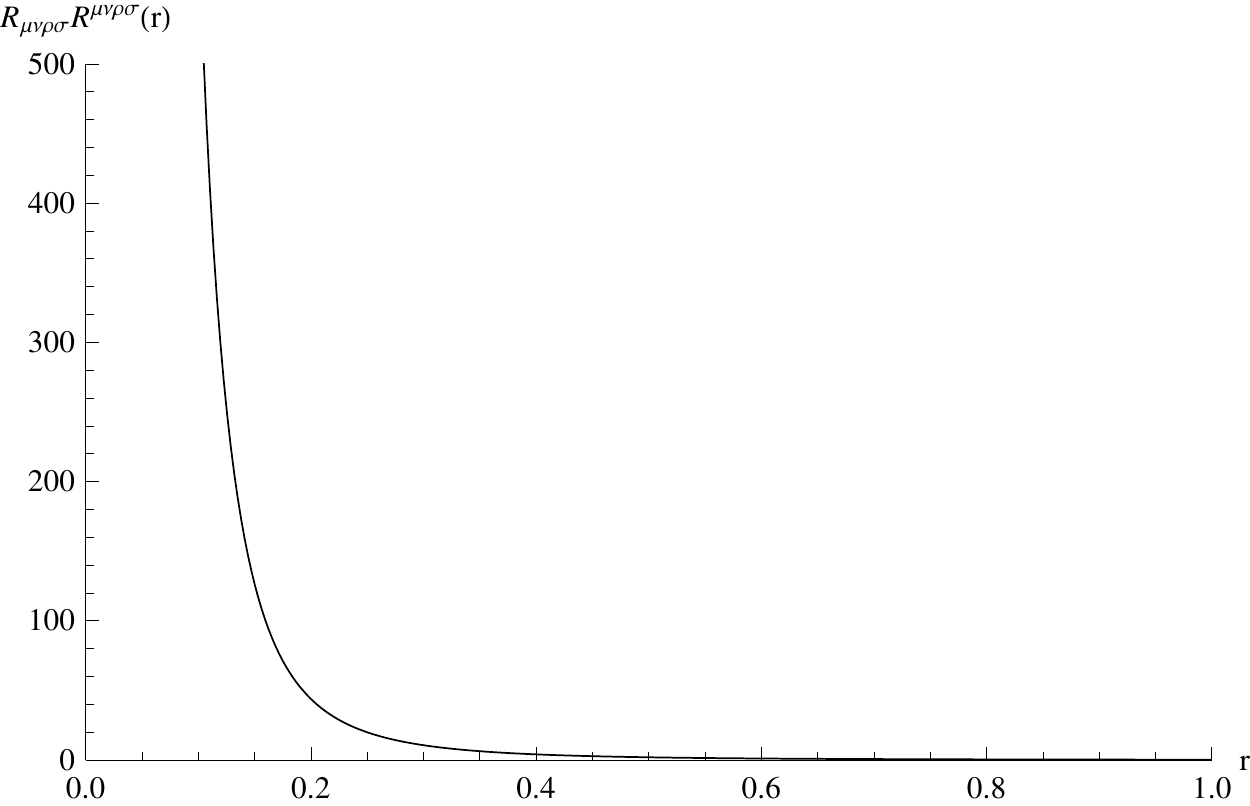}
\end{center}
\caption{The behaviour of Kretschmann scalar $R_{\mu\nu\rho\sigma}R^{\mu\nu\rho\sigma}(r)$ as function of $r$ for $\nu =3$, $C_2=1$, $q=0.1$ and $k=1$.} \label{figuraRR}
\end{figure}

Then we can investigate if our system has a hairy charged black
hole solution for $C_1=-\frac{4k}{\nu^2}$, i.e. for
$\Lambda_{eff}=0$. In Fig. \ref{plots0}  we plot the behaviour of
the metric function $f\left( r\right) $ of \ref{f(r)})
for a choice of parameters $\nu =3$,  $C_2= 1$, $10$ and $100$,
and $q=0.1$ and $k=\pm 1,0$. The metric function $f(r)$ changes
sign for low values of $r$ signalling the presence of an horizon
for $k=1$, while the potential  asymptotically vanishes
and  the scalar field is regular everywhere outside the event
horizon and null at large spatial distances as can be seen in Fig.
\ref{plots11}. Also we check the behaviour of $f(r)$ for different
values of $q$ in Fig. \ref{plotsR1}. Additionally, we have checked
the behaviour of the Kretschmann scalar
$R_{\mu\nu\rho\sigma}R^{\mu\nu\rho\sigma}(r)$ outside the black
hole horizon. As it is shown in Fig. \ref{figuraRR} there is no
curvature singularity outside the horizon for $k=1$.

Therefore for $k=1$, $\nu=3$ and a small value $q$ of the charge
of the black hole we found a well behaved charged hairy black hole
solution in asymptotically flat  spacetime. This is an unexpected result. The
presence of a cosmological constant in the gravity action
introduces a scale which protects the scalar field from getting
infinite on the horizon. All possible infinities are hidden behind
the horizon. This is the case for most of the existing hairy black
hole solutions
\cite{Martinez:2002ru,Torii:2001pg,Martinez:2004nb,Martinez:2005di,
Martinez:2006an,Kolyvaris:2009pc}. In our case the scalar field
introduces a scale by itself which makes it regular on the
horizon. Then depending on the appropriate choice of the
parameters the scale of the cosmological constant is cancelled by
the scale of the scalar field allowing in this way hairy black
holes solutions in asymptotically flat spacetime. Note that this mechanism does
not work if the charge is absent as in \cite{Gonzalez:2013aca}, because the possible examples of hairy black hole configurations violating the no-hair theorems and they were not physically acceptable as the scalar field was divergent on the horizon and stability analysis showed that they were unstable \cite{bronnikov}.

For a massless scalar field minimally coupled to gravity it was
shown in \cite{Xanthopoulos:1989kb} that no hairy black hole
solutions exist in asymptotically flat spacetimes. In our case the
scalar field is coupled minimally to gravity but it has a
non-trivial self-interaction potential. As can be seen in the
second graph of  Fig. \ref{plots11} for small values of the
parameter $\nu$ for which our system has hairy solutions, the
scalar field goes to zero very fast outside the horizon of the
black hole. Non-zero values of the scalar field are attended only
for large values of the parameter $\nu$.

A similar mechanism works in a class of Hordenski theories where a
scale is introduced in the   scalar sector and hairy black hole
solutions can be found in asymptotically flat spacetime. In these
theories there is a derivative coupling of a scalar field to
Einstein tensor. The derivative coupling has the dimension of
length square and it was shown that acts as an effective
cosmological constant \cite{Amendola:1993uh,Sushkov:2009hk}. Then
in \cite{Kolyvaris:2011fk,Kolyvaris:2013zfa}  a gravitating system
of vanishing cosmological constant consisting of an
electromagnetic field and a scalar field coupled to the Einstein
tensor was discussed. A RN black hole undergoes a second-order
phase transition  to a hairy black hole. The no-hair theorem is
evaded due to the coupling between the scalar field and the
Einstein tensor. Similar results were found in
\cite{Anabalon:2013oea}.

\begin{figure}[h]
\begin{center}
\includegraphics[scale=.7, natwidth=0.4, natheight=0.5]{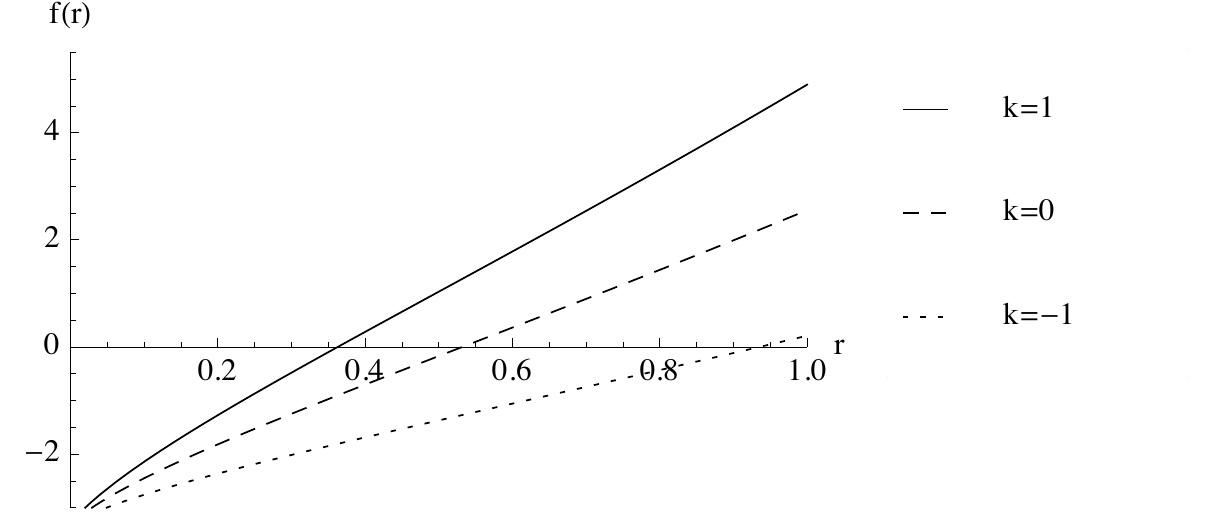}
\includegraphics[scale=.7, natwidth=0.4, natheight=0.5]{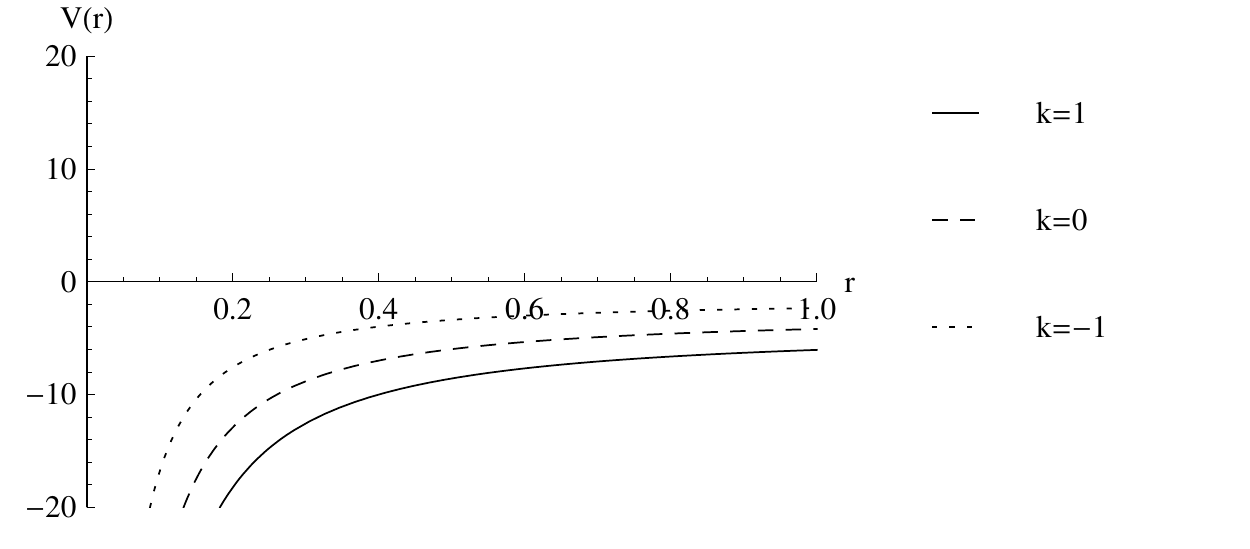}
\end{center}
\caption{The behaviour of $f(r)$ and  $V(\phi)$, for $\protect%
\nu =3$, $C_1=1$, $C_2=10$ and $q=0.1$.} \label{plots3}
\end{figure}

\begin{figure}[h]
\begin{center}
\includegraphics[scale=.7, natwidth=0.4, natheight=0.5]{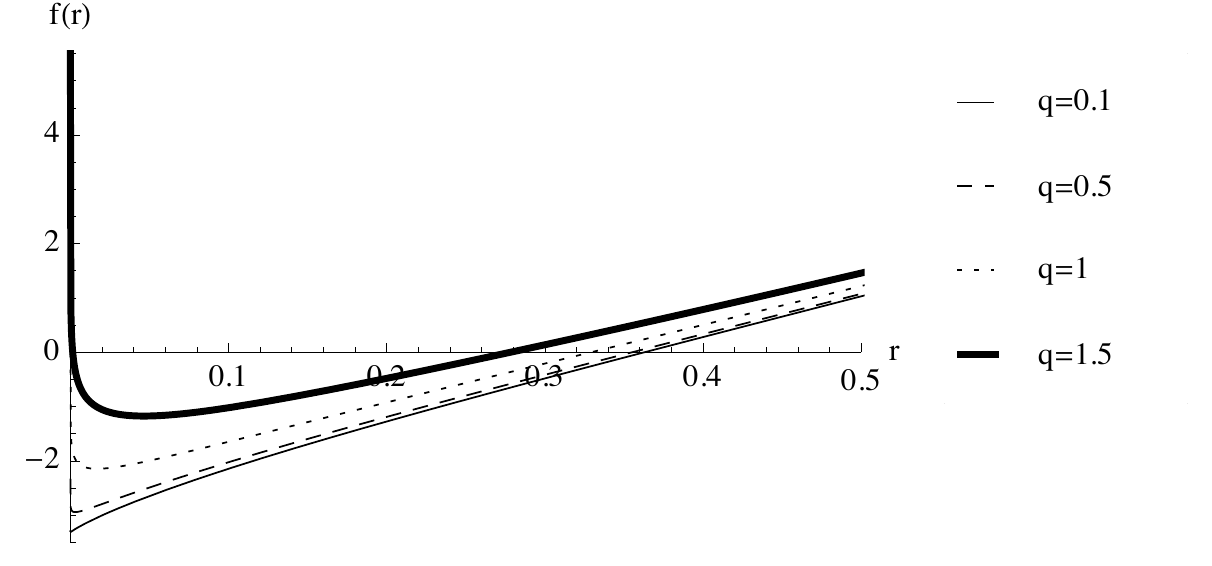}
\includegraphics[scale=.7, natwidth=0.4, natheight=0.5]{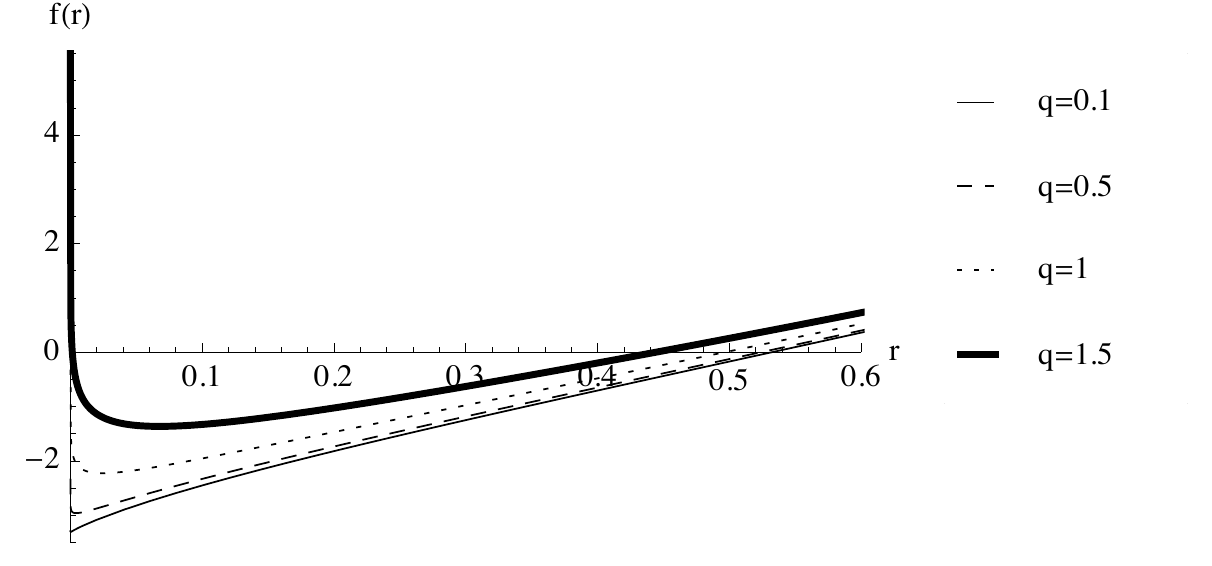}
\includegraphics[scale=.7, natwidth=0.4, natheight=0.5]{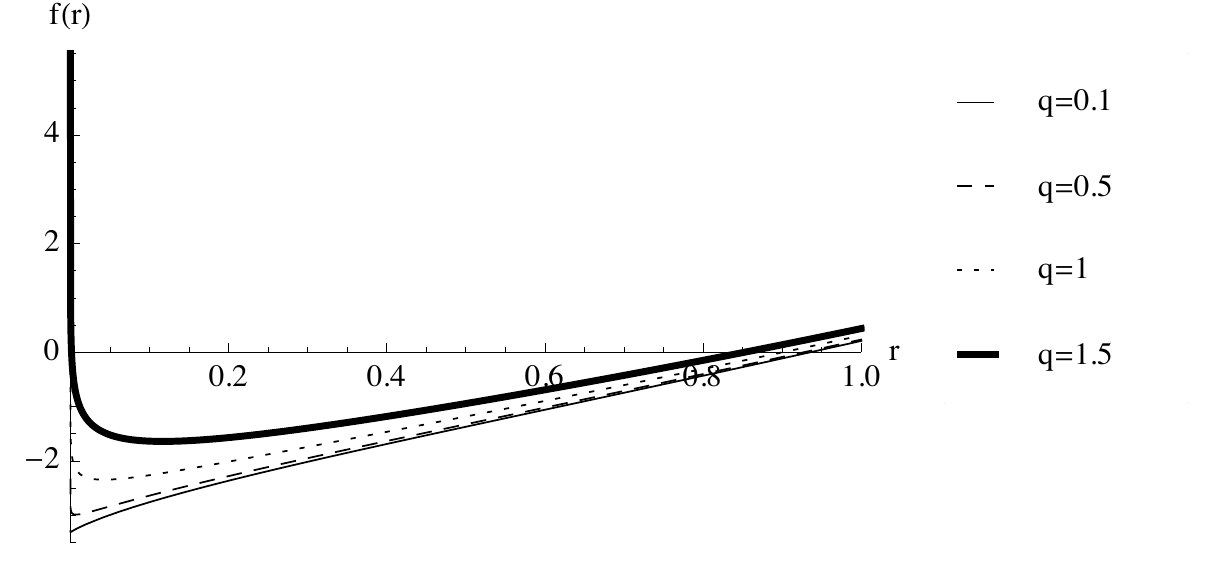}
\end{center}
\caption{The behaviour of $f(r)$, for $\protect%
\nu =3$, $C_1=C_2=1$, $k=1$ left figure, $k=0$ right figure, and $k=-1$ bottom figure for $q=0.1, 0.5,1,1.5$.} \label{plots4}
\end{figure}

\begin{figure}[h]
\begin{center}
\includegraphics[scale=.7, natwidth=0.4, natheight=0.5]{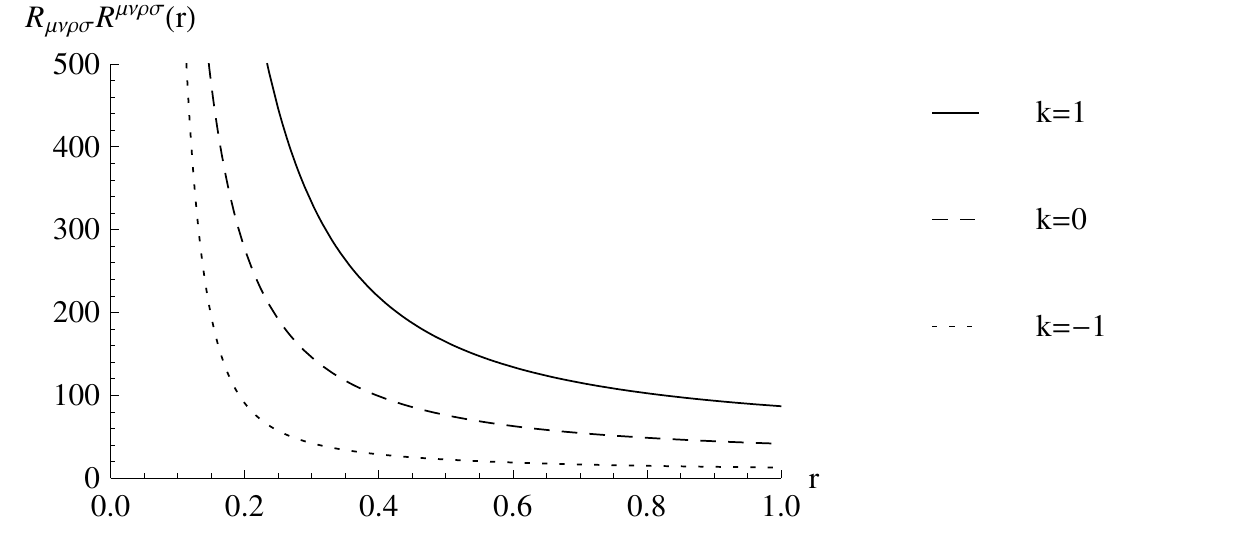}
\end{center}
\caption{The behaviour of Kretschmann scalar $R_{\mu\nu\rho\sigma}R^{\mu\nu\rho\sigma}(r)$ as function of $r$ for $\nu =3$, $C_1=C_2=1$ and $q=0.1$.} \label{figuraR}
\end{figure}

 If we allow  $\Lambda_{eff} \neq 0$ we find a general hairy
charged black hole solution and if we take the limit $q
\rightarrow 0$ we recover our previous solution found in
\cite{Gonzalez:2013aca}. Also, when $\nu \rightarrow 0$ we recover
the RN-AdS black hole. Indeed from (\ref{f(r)}) if
$C_1=-\frac{4k}{\nu^2}-\frac{\Lambda_{eff}}{3}$ the function
$f(r)$ can be written as
\begin{eqnarray}
 f\left( r\right) &= &-2\frac{q^2}{\nu^2}-(\frac{4k}{\nu^2}+\frac{\Lambda_{eff}}{3})r(r+\nu)-
\frac{C_2(2r+\nu)}{\nu^2}+2\frac{k r(2r+\nu)} {\nu^2} \nonumber
\\
&& -2\left(\frac{ q^2(2r+\nu)+r(r+\nu)(C_2+k\nu)
}{\nu^3}+\frac{q^2r(r+\nu)ln\frac{r}{r+\nu}}{\nu^4}\right)
ln\frac{r}{r+\nu}~, \label{f(rr)}
\end{eqnarray}
and in the limit  $\nu \rightarrow 0$ we recover the RN AdS black hole
\begin{equation}\label{RNL2}
f(r)=k-\frac{\Lambda_{eff}}{3}r^2+\frac{q^2}{2r^2}-\frac{C_2}{3r}~.
\end{equation}
In Fig. \ref{plots3} we plot the  behaviour of the metric function
$f\left( r\right) $ of (\ref{f(rr)}) and the potential $V\left(
r\right) $ for a choice of parameters $\nu =3$, $C_1=1$, $C_2=10$
and $q=0.1$ and $k=\pm 1, 0$. The metric function $f(r)$ changes
sign for low values of $r$ signalling the presence of an horizon,
while the potential asymptotically tends to a negative constant
(the effective cosmological constant), and  the scalar field is
regular everywhere outside the event horizon and null at large
distances. Also we consider the metric function for different
values of $q$ in Fig. \ref{plots4}. We have also checked the
behaviour of the Kretschmann scalar
$R_{\mu\nu\rho\sigma}R^{\mu\nu\rho\sigma}(r)$ outside the black
hole horizon. As it is shown in Fig. \ref{figuraR} there is no
curvature singularity outside the horizon for $k = \pm 1, 0$.

\section{Near-horizon geometry for extremal  hairy black hole}
\label{secs.4}

\begin{figure}[h]
\begin{center}
\includegraphics[scale=.7, natwidth=0.4, natheight=0.5]{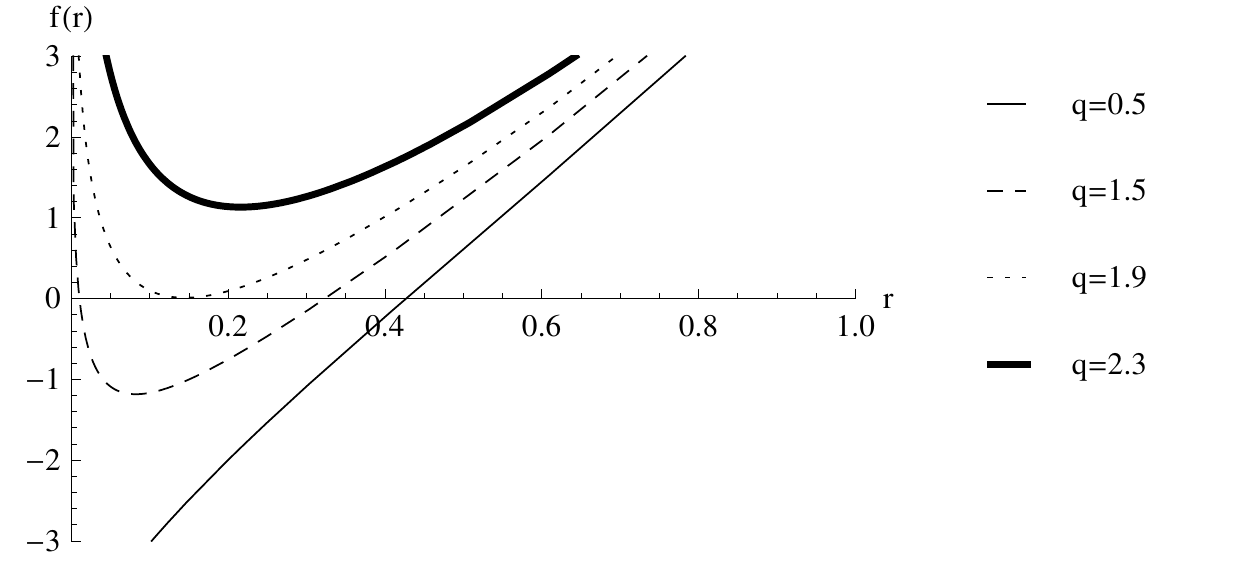}
\includegraphics[scale=.7, natwidth=0.4, natheight=0.5]{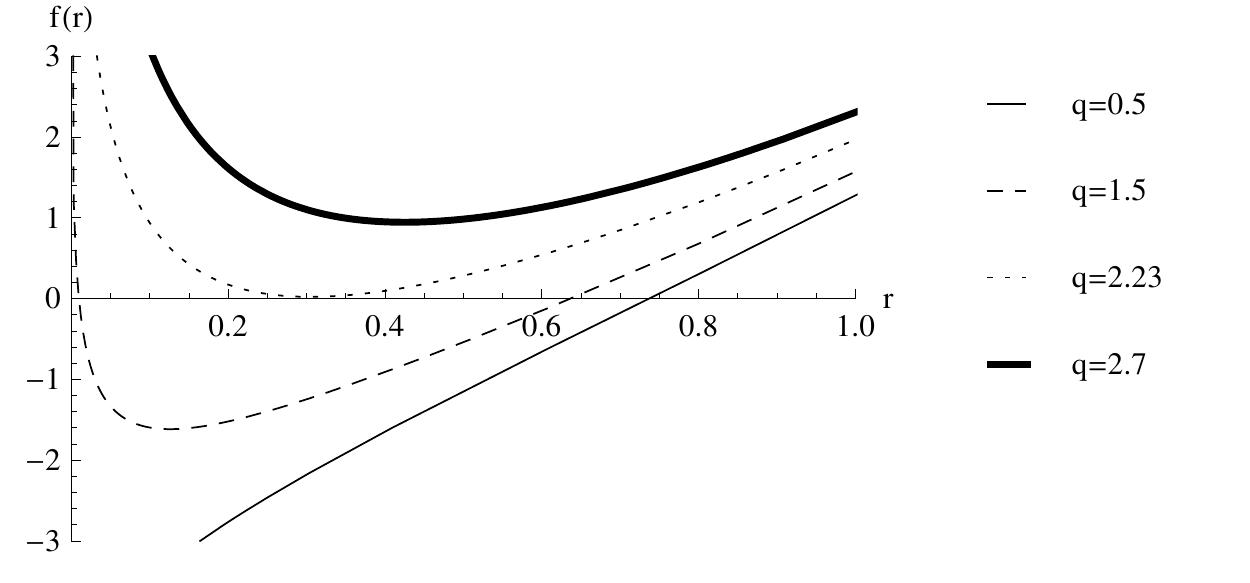}
\includegraphics[scale=.7, natwidth=0.4, natheight=0.5]{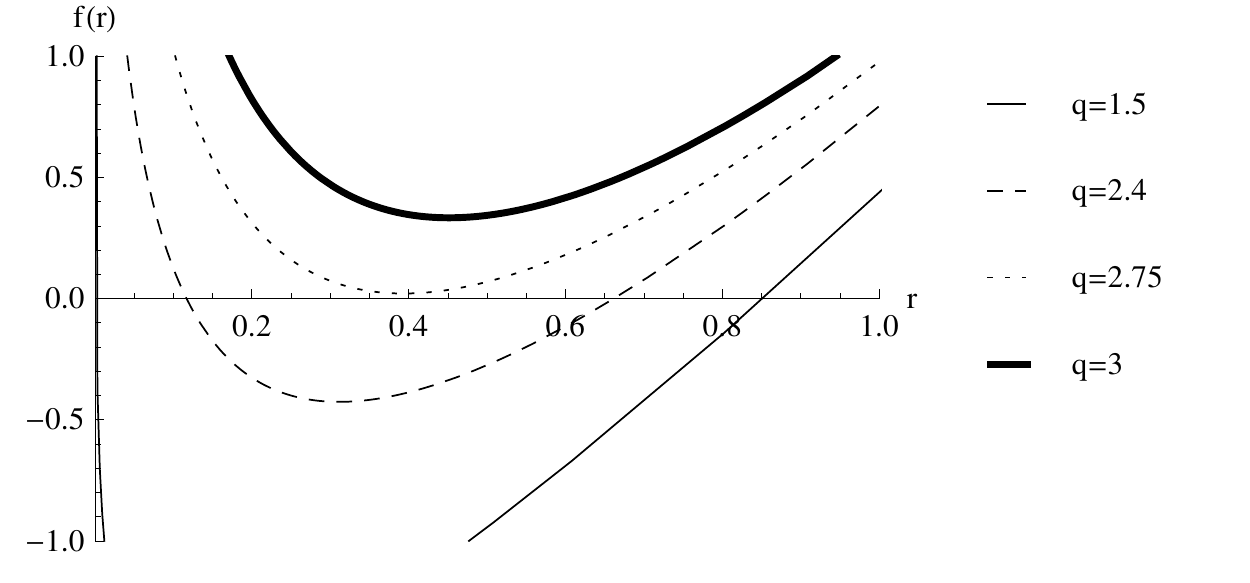}
\end{center}
\caption{The behaviour of the lapse function $f(r)$ for different
values of $q$,  $C_1=1$, $C_2=10$. Left figure for $k=1$ and
$\nu=2$. Right figure for $k=0$ and $\nu=2$, and bottom figure for
$k=-1$ and $\nu=3$} \label{figuraq}
\end{figure}

\begin{figure}[h]
\begin{center}
\includegraphics[scale=.7, natwidth=0.4, natheight=0.5]{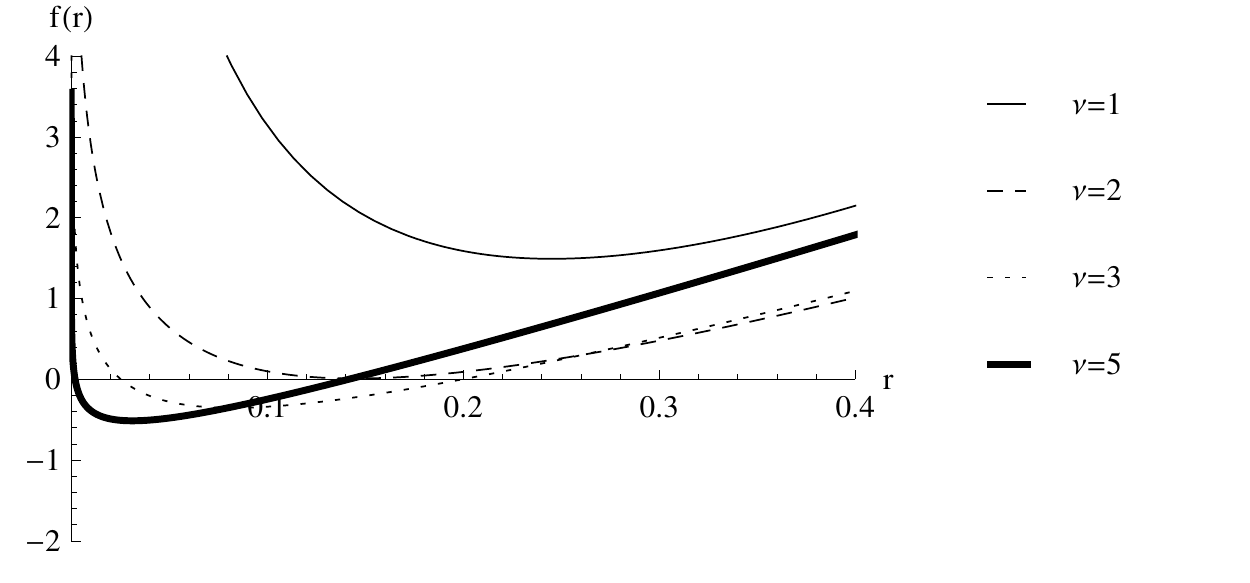}
\includegraphics[scale=.7, natwidth=0.4, natheight=0.5]{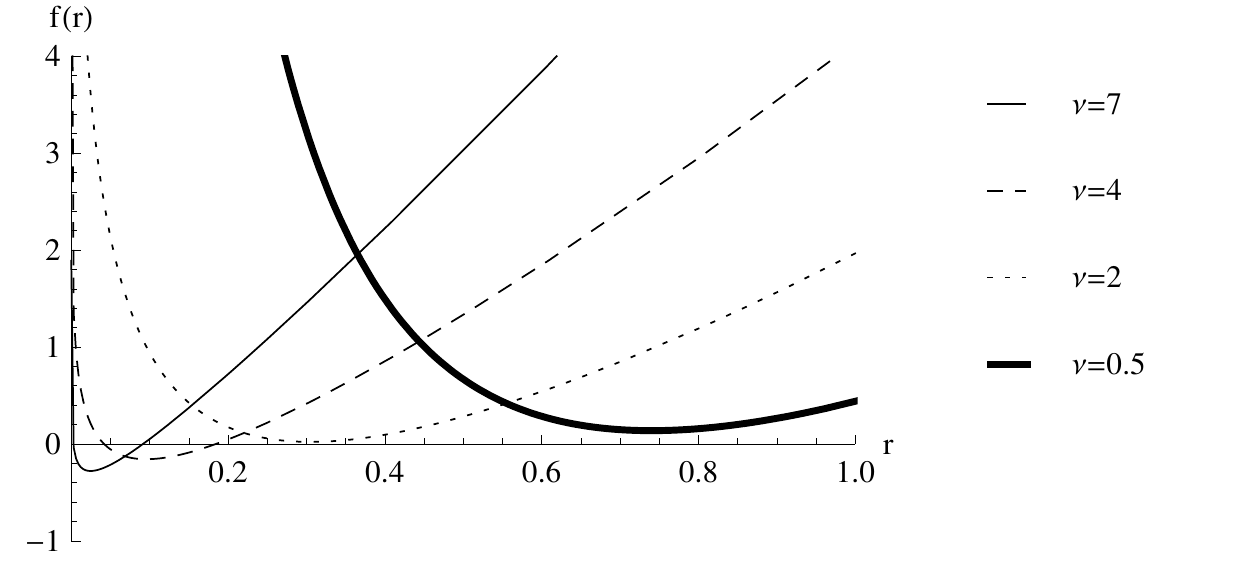}
\includegraphics[scale=.7, natwidth=0.4, natheight=0.5]{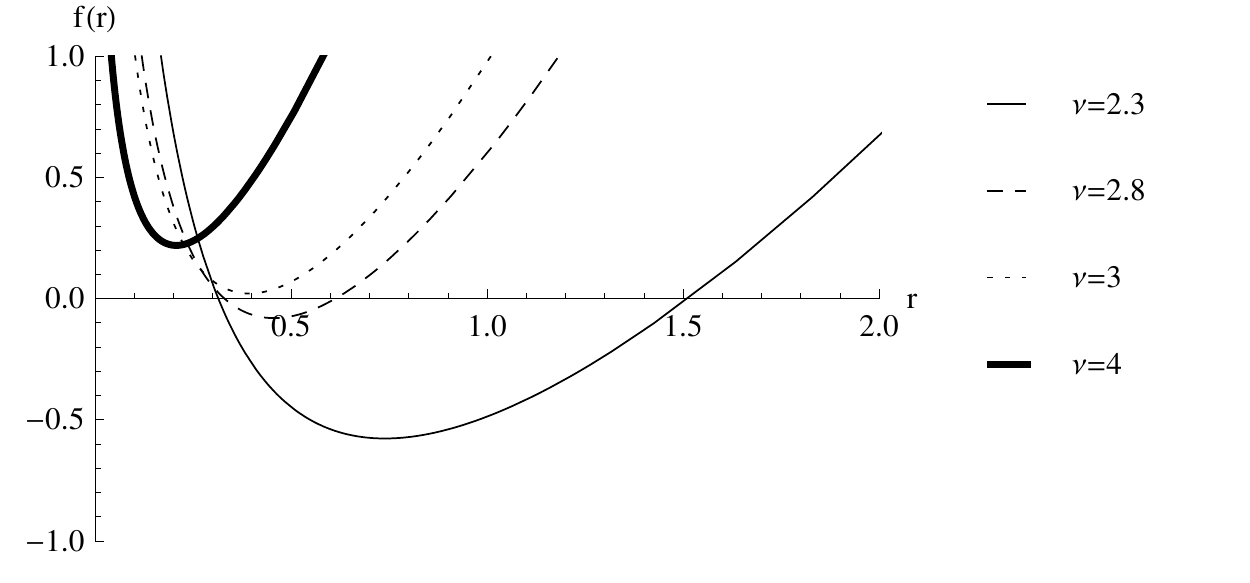}
\end{center}
\caption{The behaviour of the lapse function $f(r)$ for different
values of $\nu$,  $C_1=1$, $C_2=10$. Left figure for $k=1$ and
$q=1.9$. Right figure for $k=0$ and $q=2.23$, and bottom figure
for $k=-1$ and $q=2.75$.} \label{figuran}
\end{figure}

In this section we will investigate what is the effect of the
scalar field to the near horizon geometry of the hairy black hole
solutions we discussed in the previous section as the temperature
goes to zero. For the case of $\Lambda_{eff} \neq 0$  of an AdS
hairy black hole solution the temperature is
\begin{equation}\label{TS}
T=\frac{f^{\prime}(r_H)}{4\pi}=\frac{1}{4\pi}\frac{-2q^2\nu^2+C_1r_H\nu^4(r_H+\nu)
-2r_H(r_H+\nu)ln(\frac{r_H}{r_H+\nu})(k\nu(2r_H+\nu)-q^2ln(\frac{r_H}{r_H+\nu}))}{r_H\nu(r_H
+\nu)(\nu(2r_H+\nu)+2r_H(r_H+\nu)ln(\frac{r_H}{r_H+\nu}))}~.
\end{equation}
 Note that
$q$ has dimension of $[L]$ and it is convenient to parametrize it
as
\begin{equation}
q=\sqrt{\frac{\nu
r_*(r_*+\nu)(-C_1\nu^3+2k(2r_*+\nu)ln(\frac{r_*}{r_*+\nu}))}{2(-\nu^2+r_*(r_*+\nu)(ln(\frac{r_*}{r_*+\nu}))^2)}}~,
\end{equation}
where $r_*$ is a length scale. So in the zero temperature limit
$T=0\rightarrow r_H=r_*$ and the constant $C_2$ goes to
\begin{equation}
C_2=\frac{-2k\nu^2r_*+r_*(r_*+\nu)ln(\frac{r_*}{r_*+\nu})(C_1\nu^3-2k(r_*+\nu)ln(\frac{r_*}{r_*+\nu}))}{-\nu^2+r_*(r_*+\nu)(ln(\frac{r_*}{r_*+\nu}))^2}~.
\end{equation}
 Also, in this limit the lapse function develops a double zero at the horizon
 \begin{equation}
 f(r)=\eta (r-r_*)^2+...~,
 \end{equation}
 where $\eta$ is given by
 \begin{equation}
\eta=
\frac{2k\nu^2-C_1\nu^4-2k(-\nu(2r_*+\nu)+r_*(r_*+\nu)ln(\frac{r_*}{r_*+\nu}))ln(\frac{r_*}{r_*+\nu})}{2r_*(r_*+\nu)(-\nu^2+r_*(r_*+\nu)(ln(\frac{r_*}{r_*+\nu}))^2)}~.
 \end{equation}
 Now, considering the scaling limit
\begin{equation}
r-r_*=\lambda\frac{R^2}{\chi}~,~
t=\lambda^{-1}\tau~,~\lambda\rightarrow 0~,
\end{equation}
with $\chi$ and $\tau$ finite, where
\begin{equation}
R=\frac{1}{\sqrt{\eta}}~,
\end{equation}
we find that the metric (\ref{metricBH}) can be written
 as $AdS_2~\times~\Omega_2$ \cite{Kunduri:2013gce}, where $ \Omega_2$ is the
base manifold
\begin{equation}
ds^2=\frac{R^2}{\chi^2}(-d\tau^2+d\chi^2)+r_*(r_*+\nu)d\Omega_2~.
\end{equation}
Note that the scale of the $AdS_2$ space $R$ depends on the charge
of the scalar field $\nu$. In Fig. (\ref{figuraq}) we show the
behaviour of the lapse function $f(r)$ for different values of $q$
for fixed $\nu$. We see that for each value of $k$ there  is
critical value of $q_c$ above which the extremal black hole is
destabilized while below that value we depart from the extremal
limit and we go to the charged hairy black hole solution. Similar
behaviour we observe for the charge $\nu$ in Fig. (\ref{figuran}),
where we show the behaviour of the lapse function $f(r)$ for
different values of $\nu$ for fixed $q$. Observe here that the
behaviour of the lapse function $f(r)$ is more sensitive to $k$
than in the $q$ case. In summary, there is a pair of critical
values of ($q_c,\nu_{c}$) where the extremal black hole is formed
and the near horizon geometry is given by $AdS_2~\times~\Omega_2$.
Then there is a range of values of ($q$, $\nu$) which lead to AdS
space or to no-extremal hairy solutions.


\begin{figure}[h]
\begin{center}
\includegraphics[scale=.7, natwidth=0.4, natheight=0.5]{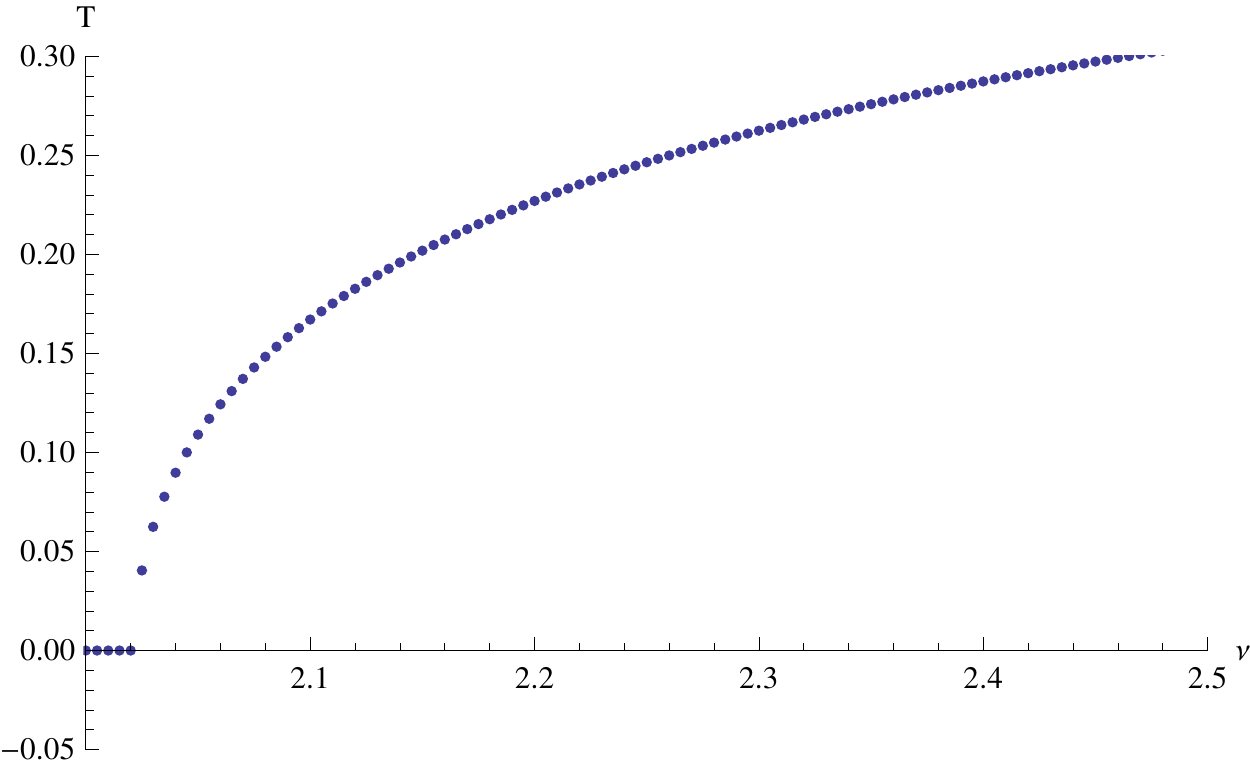}
\includegraphics[scale=.7, natwidth=0.4, natheight=0.5]{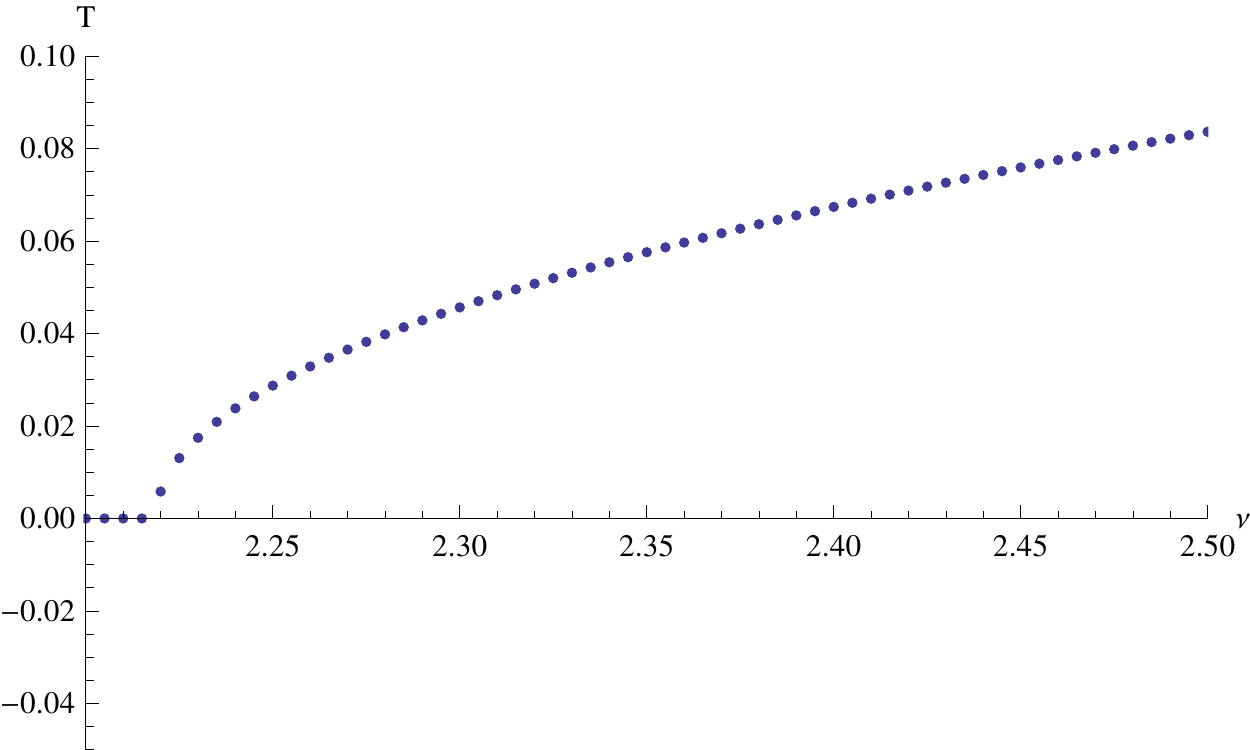}
\includegraphics[scale=.7, natwidth=0.4, natheight=0.5]{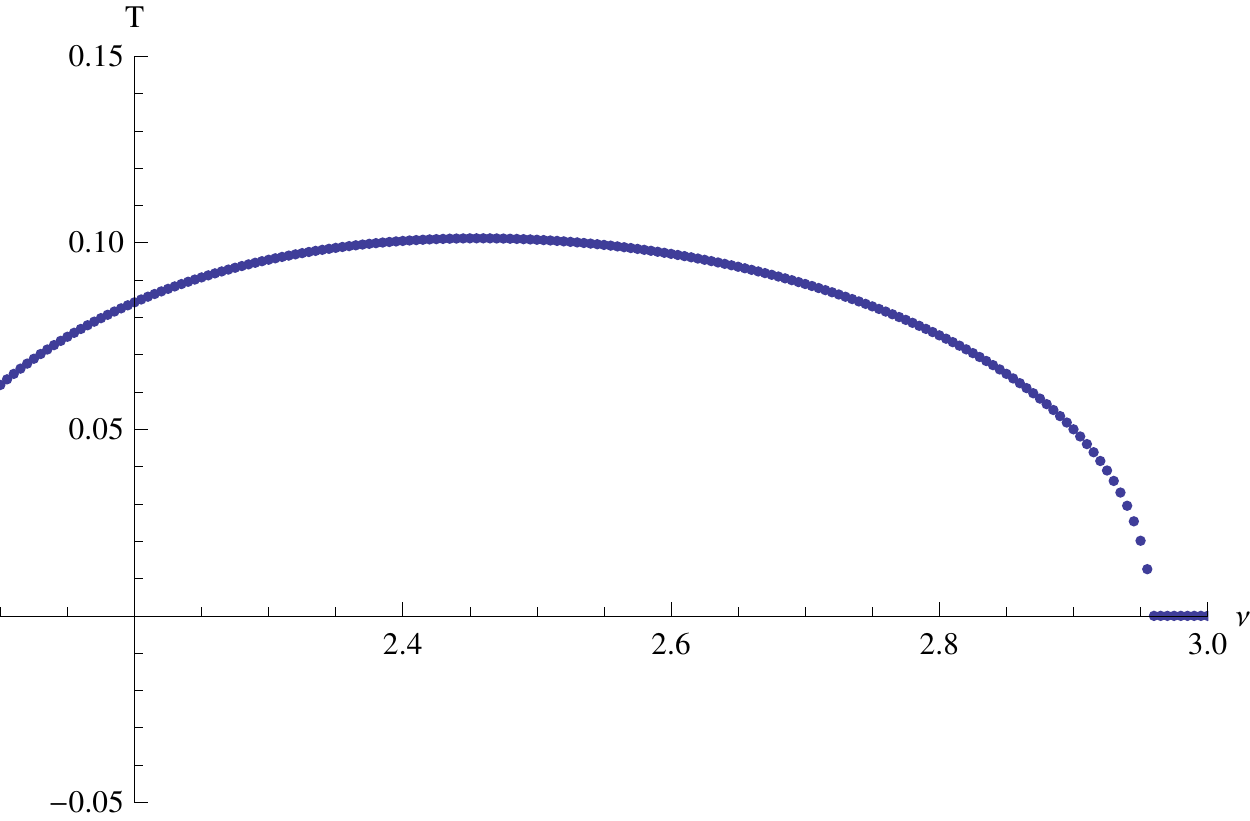}
\end{center}
\caption{The behaviour of the temperature $T$ as a function of
$\nu$,  $C_1=1$, $C_2=10$. Left figure for $k=1$ and $q=1.9$.
Right figure for $k=0$ and $q=2.23$, and bottom figure for $k=-1$
and $q=2.75$.} \label{figuraT}
\end{figure}

It is interesting to see how the temperature changes for various
values of $\nu$ and for the fixed value of $q=q_c$ of the
corresponding value of $k$. In Fig. \ref{figuraT} we see that
there is a range of values of $\nu$ in which the temperature
remains zero. Then above a critical value $q_c$ for $k=1,0$ the
hairy black hole is thermalized while for $k=-1$ as $\nu$ is lower
below $\nu_c$ the hairy black hole researches its maximum
temperature and then starts to cool down. This behaviour is
interesting and it deserves to be studied further.

\begin{figure}[h]
\begin{center}
\includegraphics[scale=.7, natwidth=0.4, natheight=0.5]{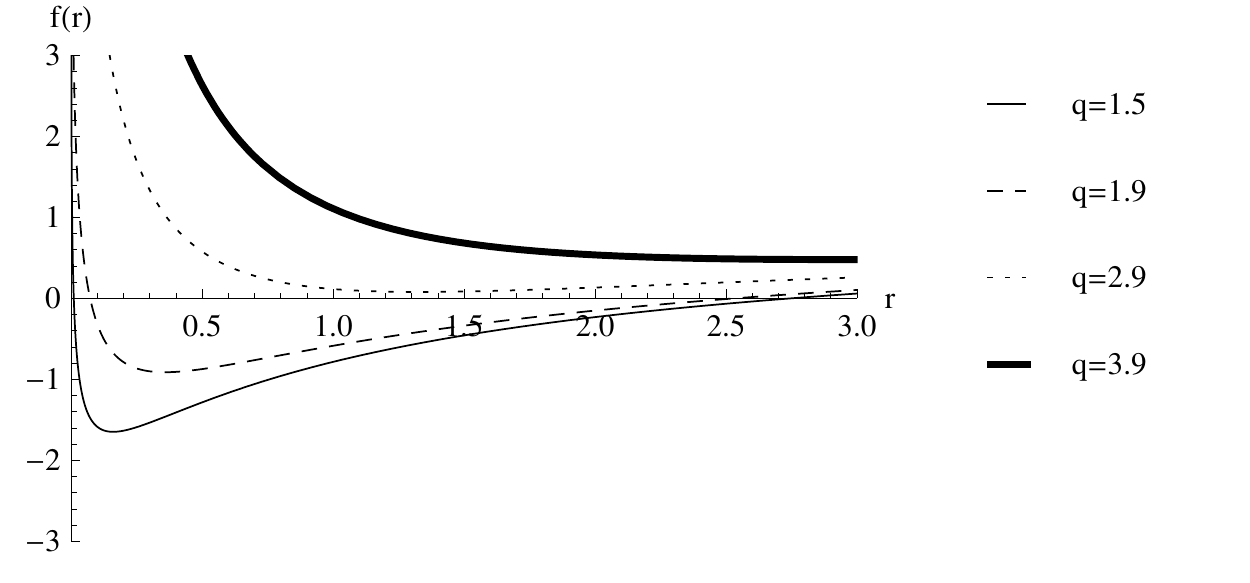}
\includegraphics[scale=.7, natwidth=0.4, natheight=0.5]{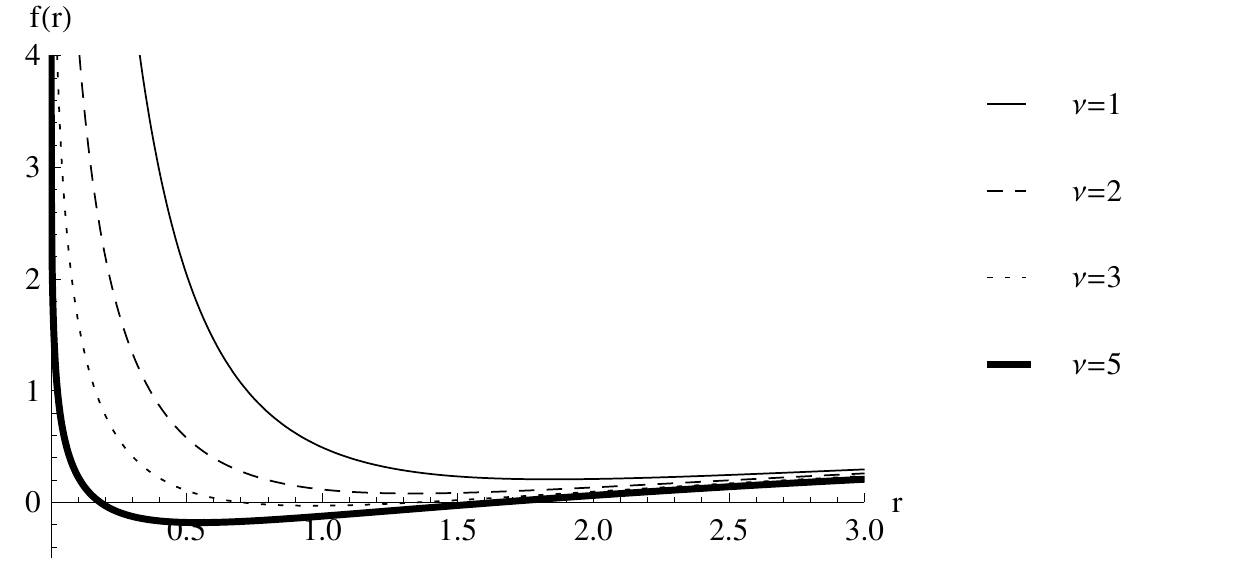}
\includegraphics[scale=.7, natwidth=0.4, natheight=0.5]{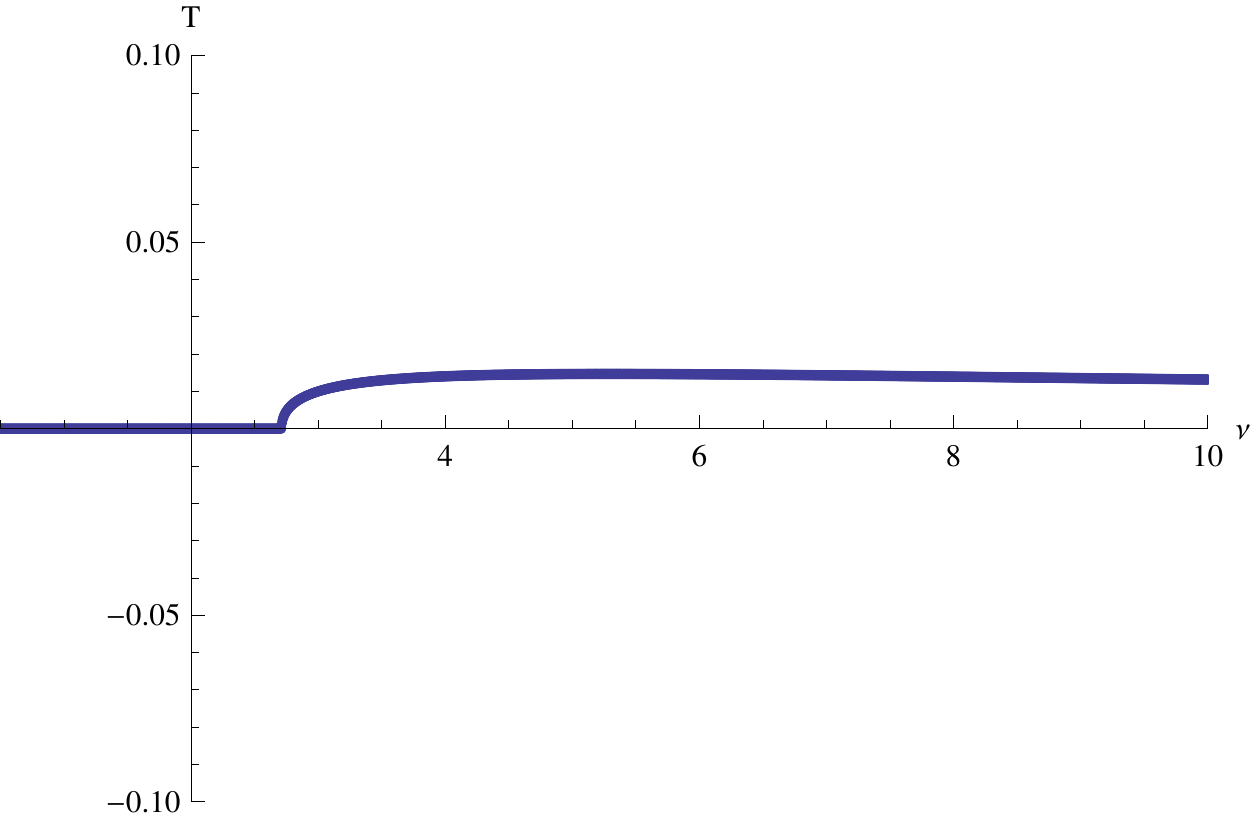}
\end{center}
\caption{The behaviour of the lapse function $f(r)$ for different
values of $q$, $k=1$, $C_2=10$, and $\nu = 2$ (left figure). The behaviour of the lapse function $f(r)$ for different
values of $\nu$, $k=1$, $C_2=10$, and $q=2.9$ (right figure). The behaviour of the temperature $T$ as a function of
$\nu$, $k=1$, $C_2=10$, and $q=2.9$ (bottom figure).} \label{figuraTC1}
\end{figure}

In the case of $\Lambda_{eff}=0$ i.e. $C_1=-\frac{4k}{\nu^2}$, we
 observe a similar behaviour. In this case a hairy black hole solution exist only for $k=1$ (see Fig. \ref{plots0}).
 Then in Fig. \ref{figuraTC1} we can see that  there is
a critical value of $q_c$ above which the extremal black hole is
destabilized while below that value we depart from the extremal
limit and we go to the charged hairy black hole solution. However
in this case the extremal double horizon is larger than in the
case of $\Lambda_{eff}\neq 0$. Also, the same behaviour is
observed for $\nu$ so   there is a pair of critical values of
($q_c,\nu_{c}$) where the extremal black hole is formed and the
near horizon geometry is given by $AdS_2~\times~\Omega_2$. Then
there is a range of values of ($q$, $\nu$) which lead to flat
space or to no-extremal hairy solutions. Also, we see that there
is a range of values of $\nu$ in which the temperature remains
zero, and  a critical value $\nu_c$ above which the hairy black
hole is thermalized. Observe that in this case the temperature
remains constant above $\nu_c$.

\section{Thermodynamics}
\label{secs.5}

In this section we will study the thermodynamics of the found
hairy black hole solutions.
To apply the Euclidean formalism we will work in the $\rho
=\sqrt{r\left( r+\nu \right) }$ coordinates in which the metric
(\ref{metricnew}) can be written in the following  form
\begin{equation}
ds^{2}=N^{2}\left( \rho\right) g^{2}\left( \rho\right) d\tau ^{2}+g^{-2}\left(
\rho\right) d\rho^{2}+\rho^{2} d\sigma ^{2}~,
\end{equation}
where
\begin{eqnarray}
N\left( \rho\right)^2 =\frac{\rho ^{2}}{\left( \frac{\nu^2}{4}+\rho^2\right) }~,\,\,
g^{2}\left( \rho\right) = \frac{\chi(\rho)}{\rho^2}\left( \frac{\nu^2}{4}+\rho^2\right)~.\,\,
\end{eqnarray}
Now, we go to Euclidean time $t \rightarrow it$ and we consider the
action
\begin{equation}\label{EA}
I=-\frac{\beta\sigma}{4\pi}\int_{\rho_H}^{\infty} \left( N(\rho)\mathcal{H}(\rho)+A_{t}p^\prime \right)d\rho+B_{surf}~,
\end{equation}
where $\mathcal{H}(\rho)$ is the reduced Hamiltonian which
satisfies the constraint  $\mathcal{H}(\rho)=0$ and
$p(\rho)=\frac{\rho^2}{N(\rho)}A^\prime_t$, $p^\prime(\rho)=0$.
Also  $B_{surf}$ is a surface term, $\beta=1/T$ is the period of
Euclidean time and finally $\sigma $ is the area of the spatial
2-section. We now compute the action when the field equations
hold. The condition that the geometries which are permitted should
not have conical singularities at the horizon imposes
\begin{equation}\label{T}
T=\frac{N(\rho_H)g^2(\rho_H)^{\prime}}{4\pi }~.
\end{equation}
So, by using the grand canonical ensemble we can fix  the
temperature and "voltage" $\psi = (A_t(\infty)-A_t(\rho_H))$. Then
the variation of the surface term yields
\begin{equation}
\delta B_{surf}=\delta B_{\phi }+\delta B_{G}+\delta B_{F}~,
\end{equation}
where
\begin{equation}
\delta B_{G}=\beta \sigma \left[N\rho\delta g^2\right] _{\rho_H}^{\infty }~,
\end{equation}
\begin{equation}
\delta B_{\phi }=\beta \sigma \left[N\rho^{2}g^{2}\phi^{\prime }\delta \phi\right] _{\rho_H}^{\infty }~,
\end{equation}
\begin{equation}
\delta B_{F }=\beta \sigma \left[ A_t \delta p \right] _{\rho_H}^{\infty }~.
\end{equation}
Now, we will apply the above formalism to the cases $\Lambda_{eff}=0$
and $\Lambda_{eff}\neq 0$. In both cases the variation of the fields at large distances yields
\begin{equation}
\delta B_{G\infty }=\beta \sigma
\left[\left(-\frac{1}{3}+\mathcal{O}\left(\frac{1}{\rho^2}\right)\right)\delta
C_2+\left(
-\frac{k}{3}+\mathcal{O}\left(\frac{1}{\rho}\right)\right)\delta
\nu+\mathcal{O}\left(\frac{1}{\rho}\right)\delta q^2\right]~,
\end{equation}
\begin{equation}
\delta B_{\phi \infty }=\mathcal{O}\left(\frac{1}{\rho}\right)
\delta\nu~,
\end{equation}
and
\begin{equation}
\delta B_{F\infty}=\mathcal{O}\left(\frac{1}{\rho^2}\right)\delta
q^2+\mathcal{O}\left(\frac{1}{\rho^4}\right)\delta\nu~.
\end{equation}
On the other hand,
\begin{equation}
\delta B_{\rho_H}=\sigma\left[ -\frac{1}{2}N(\rho_H)\beta
(g(\rho_H)^2)^\prime\delta\rho_H^2+\beta \psi \delta{p}\right] ~.
\end{equation}
Thus, from the above expressions in both cases we deduce the
surface terms at large distances
\begin{equation}
B_{surf \infty }=-\frac{\beta \sigma}{3}\left ( C_2+k \nu \right)~,
\end{equation}
and at the horizon
\begin{equation}
B_{surf \rho_H}=-2\pi\sigma\rho_H^2+\frac{\beta\psi\sigma q}{2}~.
\end{equation}
Therefore,  the Euclidean action reads
\begin{equation}
I=-\frac{\beta \sigma}{3}\left ( C_2+k \nu \right)+2\pi\sigma\rho_H^2-\frac{\beta\psi\sigma q}{2}~,
\end{equation}
and as the Euclidean action is related to the free energy through $I=-\beta F$, we obtain
\begin{equation}
I=S-\beta \mathcal{M}+\beta\psi \mathcal{Q}~,
\end{equation}
where the mass $\mathcal{M}$ is
\begin{equation}\label{M}
\mathcal{M}=\frac{\sigma}{3}\left ( C_2+k \nu \right)~,
\end{equation}
the entropy $S$ is
\begin{equation}\label{S}
S=2\pi\sigma\rho_H^2~,
\end{equation}
and the electric charge $\mathcal{Q}$ is
\begin{equation}\label{Q}
\mathcal{Q}=-\frac{\sigma q }{2}~.
\end{equation}

\begin{figure}[h]
\begin{center}
\includegraphics[scale=.7, natwidth=0.4, natheight=0.5]{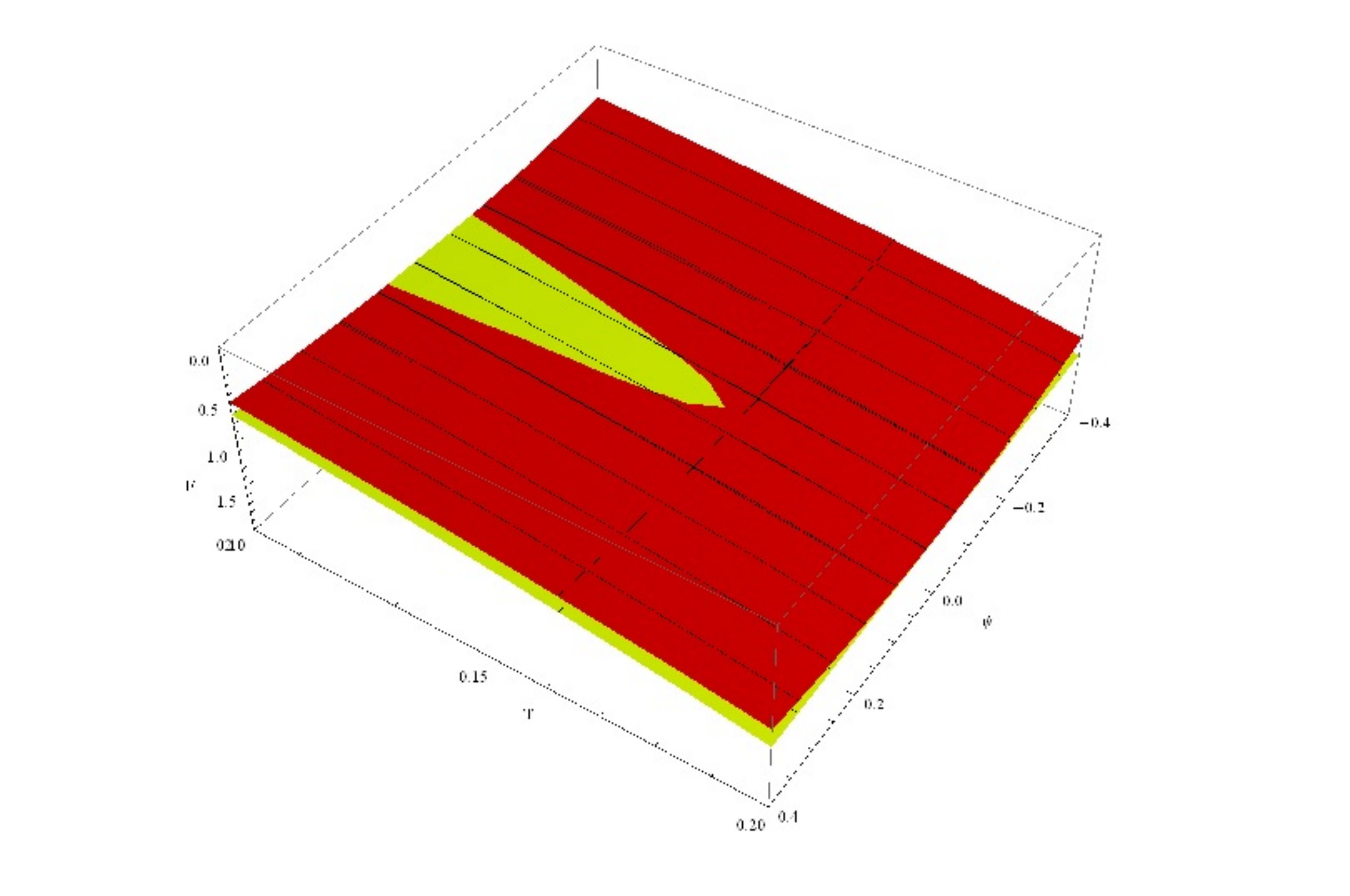}
\end{center}
\caption{The behaviour of the free energy for the charged hairy
black hole and  the RN anti-de Sitter black hole as function of
$T$ and $\psi$ with $k=-1$, $l=1$, $\sigma=1$, and $\nu=0.5$.
Red surface for charged hairy black hole and yellow surface for
the RN anti-de Sitter black hole.} \label{figura5}
\end{figure}

We will first discuss the case of $\Lambda_{eff} \neq 0$. The
temperature is given by the relation (\ref{TS}), while the mass,
entropy and electric charge  can be written respectively as
\begin{equation}
\mathcal{M}=\frac{\sigma}{3}\frac{\nu^2(-2q^2+r_H(C_1\nu^2(r_H+\nu)+2k(2r_H+\nu)))
-2ln(\frac{r_H}{r_H+\nu})(r_H(r_H+\nu)(k\nu^2+q^2ln(\frac{r_H}{r_H+\nu}))+\nu
q^2(2r_H+\nu))   }{\nu^2(2r_H+\nu)+2\nu
r_H(r_H+\nu)ln(\frac{r_H}{r_H+\nu})}+\frac{\sigma}{3}k \nu~,
\end{equation}
\begin{equation}
S=2\pi\sigma r_H(r_H+\nu)~,
\end{equation}
\begin{equation}
\mathcal{Q}=\frac{\sigma}{2}\frac{\nu\psi}{ln(\frac{r_H}{r_H+\nu})}~.
\end{equation}
We will  study possible phase transitions of the hairy black hole
solutions to known black hole solutions without hair. For
$\Lambda_{eff} \neq 0$ and in the absence of a scalar field the
action (\ref{action}) has as a solution the RN-AdS black hole
with temperature,
entropy, mass and electric charge  given respectively by
\begin{equation}
T_{RN}=\frac{1}{4\pi}(\frac{3\rho_+}{l^2}+\frac{k}{\rho_+}-\frac{\psi^2}{8\pi
\rho_+}), \text{ \ }S_{RN}=2\pi \sigma \rho _{+}^{2},\text{ \
}M_{RN}=\sigma \rho_+(k+\frac{\rho_+^2}{l^2}+\frac{\psi^2}{8\pi}),
\text{ \ }Q_{RN}=\frac{\sigma \rho_+ \psi}{4\pi}~.
\end{equation}
Then, the horizon radius $\rho _{+}=\frac{2\pi l^2 T}{3}\left(
1+\sqrt{1-\frac{3(k-\psi^2/(8\pi))}{4\pi^2 l^2 T^2}}\right)$ can
be written as a function of the temperature and of the electric
potential. Now in order to find phase transitions between the
charged hairy and RN-AdS black hole, we must consider both black
holes in a same grand canonical ensemble, i.e. at the same
temperature $T$ and electric potential $\psi$. Equaling $T$ and
$\psi$ for both black holes and by considering the free energy $F$
\begin{equation}
F =F(T,\phi)=M-TS-\psi Q~,
\end{equation}
we plot the free energy $F_0$ for the charged black hole with
scalar hair, and $F_1$ of the RN-AdS black hole  in Fig.
\ref{figura5}, in order to see the range of values of the electric
potential $\psi$ and of the temperature $T$ of the black hole for
which the phase transitions exist.

\begin{figure}[h]
\begin{center}
\includegraphics[scale=.7, natwidth=0.4, natheight=0.5]{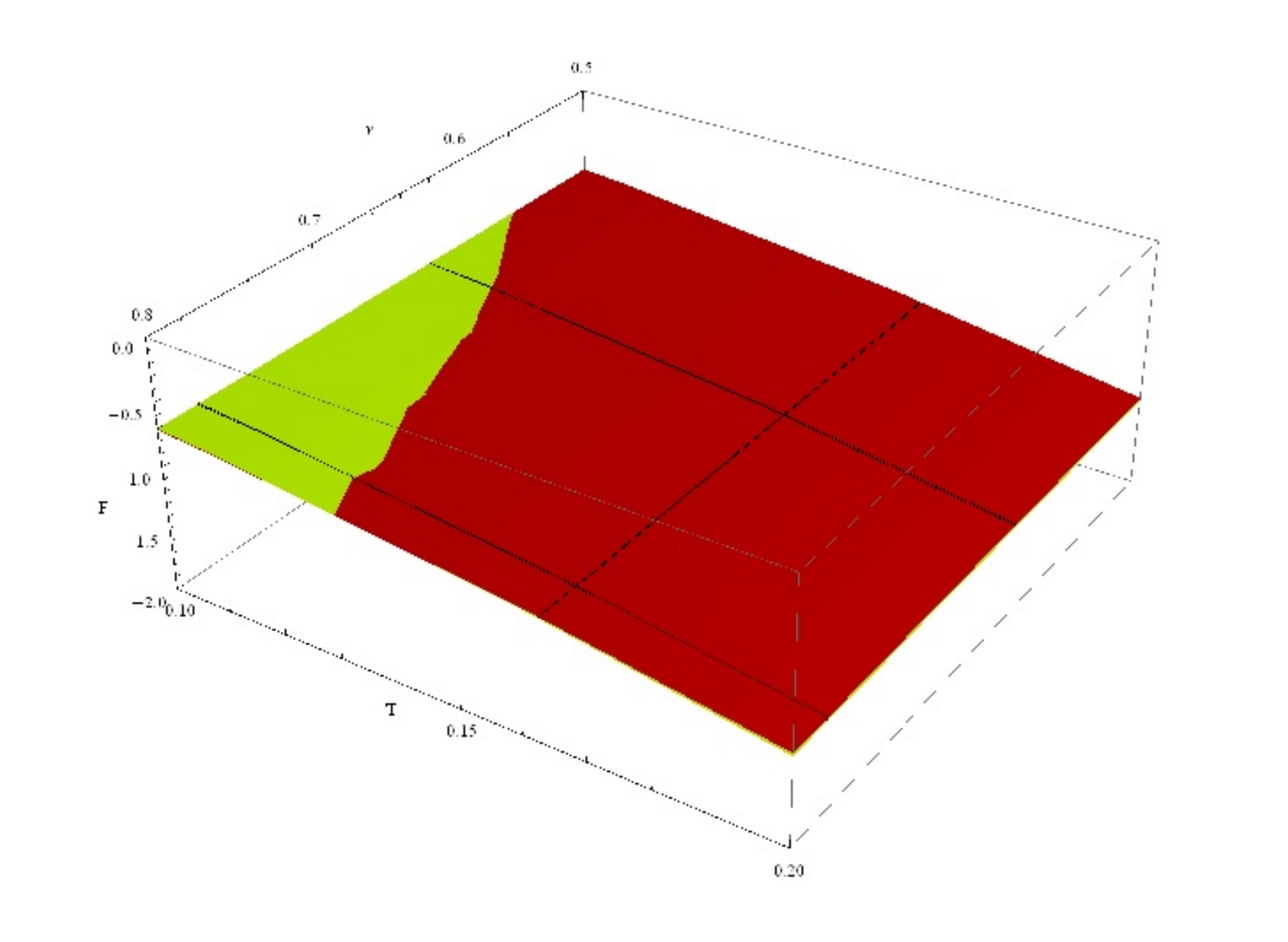}
\end{center}
\caption{The behaviour of the free energy for the charged hairy
black hole and the RN  black hole as function of $T$ and $\nu$
with $k=-1$, $\sigma=1$, $l=1$, and $\psi=0.1$. Red surface for
charged hairy black hole and yellow surface for the RN black
hole.} \label{nv1}
\end{figure}

Thus, from Fig. \ref{figura5}, we can see that there exists a
phase transition, and the charged hairy black hole dominates for
small temperatures, while for large temperatures the RN-AdS black
hole would be preferred. Also, we can observe that the critical
temperature at which this phase transition takes place depends on
the $\psi$ and as it can be seen in Fig. \ref{nv1} it depends also
on $\nu$ the charge of the scalar field at small temperatures. At
zero temperature the RN-AdS is preferable which agrees with our
previous results. This phase transition occurs for hyperbolic
horizon $k=-1$, in agreement with the findings in
\cite{Kolyvaris:2009pc, Martinez:2010ti, Gonzalez:2013aca} where
only phase transitions of exact hairy black hole solutions to
black hole solutions with hyperbolic horizons.

\begin{figure}[h]
\begin{center}
\includegraphics[scale=.7, natwidth=0.4, natheight=0.5]{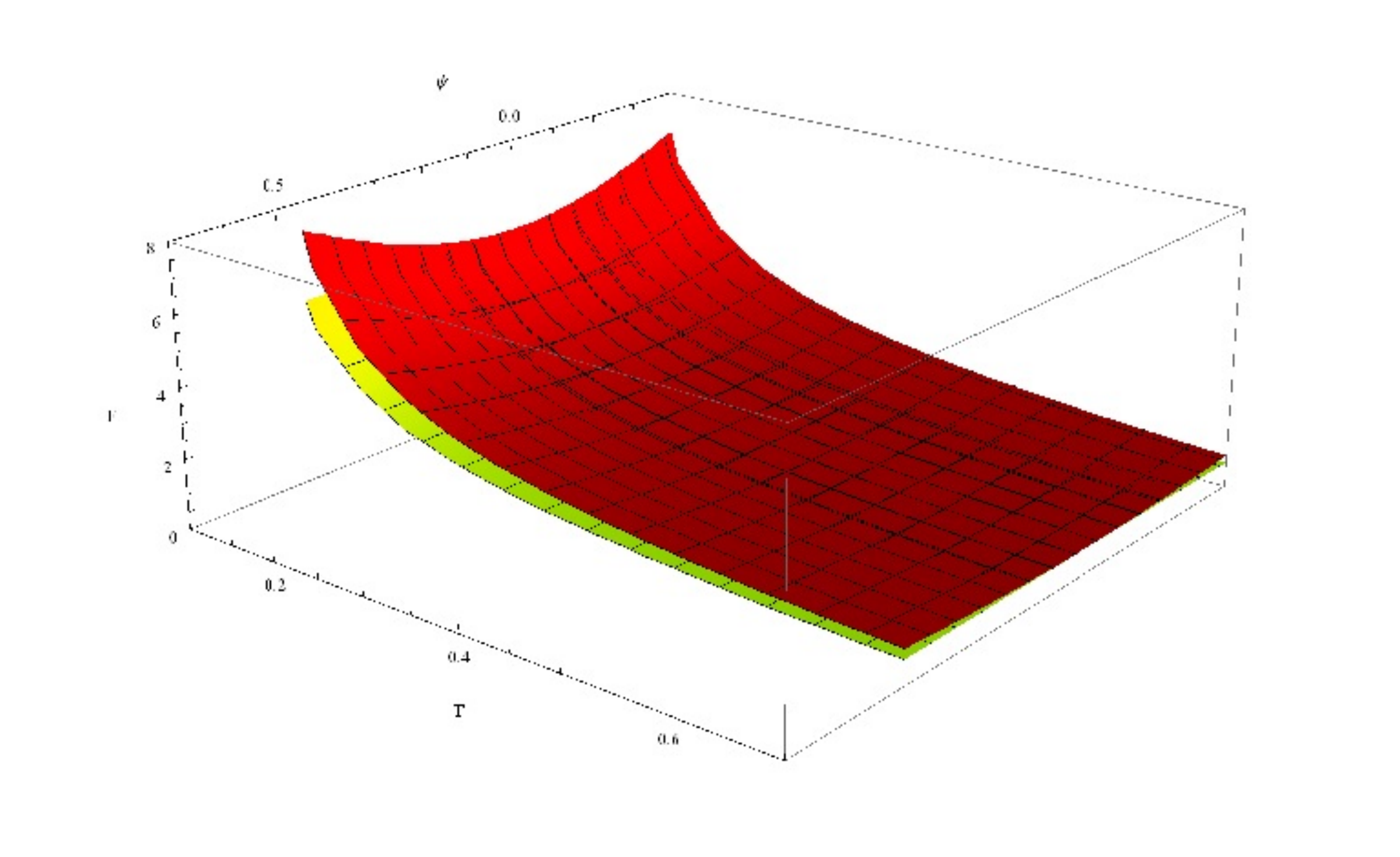}
\end{center}
\caption{The behaviour of the free energy for the charged hairy
black hole and the RN black hole as function of $T$ and $\nu$ with
$k=1$, $\sigma=4\pi$, and $\nu=0.01$. Red surface for charged hairy
black hole and yellow surface for the RN black hole,
$C_1=-\frac{4k}{\nu^2}$.} \label{figura55}
\end{figure}

A similar analysis can be carried  out for the case of
$\Lambda_{eff} = 0$.   In this case the temperature, mass, entropy
and electric charge are given respectively by
\begin{equation}
T_{RN}=\frac{1}{4\pi \rho _{+}}\left( 1-\frac{\psi^2}{8\pi}\right) ,%
\text{ \ }S_{RN}=2\pi \sigma \rho _{+}^{2},\text{ \ }M_{RN}=\sigma
\rho _{+}\left( 1 + \frac{\psi^{2}}{8\pi}\right),\text{ \
}Q_{RN}=\frac{\sigma \rho_+ \psi}{4\pi}~,
\end{equation}
and the horizon radius is given by $\rho _{+}=\frac{1}{4\pi
T_{RN}}\left( 1-\frac{\psi^2}{8\pi}\right)$. In Fig.
\ref{figura55}  we plot the free energy $F_0$ for the charged
black hole with scalar hair  and $F_1$ the free energy of the RN
black hole.
Then   we can see that there not exists a phase transition, and the RN
black hole dominates for all temperatures.



\section{Conclusions}
\label{secs.6}

We have considered a gravitating system consisting of a scalar
field minimally coupled to gravity with a self-interacting
potential and an U(1) electromagnetic field. We solved exactly the
coupled Einstein-Maxwell-scalar field equations with a profile of
the scalar field which falls sufficient fast outside the black
hole horizon.  For a range of  values of the scalar field
parameter, which characterizes its behaviour,  we found exact
hairy charged black hole solutions with the scalar field  regular
everywhere.

The presence of the scalar field introduced a scale in the system,
resulting in a redefinition of the cosmological constant to
$\Lambda_{eff}$. If $\Lambda_{eff} \neq 0 $ then hairy black hole
solutions were found for $k=\pm 1, 0$ with the potential at large
distances to tend to the effective cosmological constant, and the
scalar field to be regular everywhere outside the event horizon
and null at large distances. If $\Lambda_{eff} = 0 $ then a hairy
charged black hole solution was found for $k=1$ in flat space. In
both cases if the scalar field is decoupled then the RN black hole
solution is recovered.

Because the scalar hair is non-zero only near the horizon of the
black hole, we studied the effect of the scalar field on the near
horizon limit of external black hole as the temperature goes to
zero. We found that except a critical value of the charge of the
black hole there exist also a critical value of scalar field
beyond which the extremal black hole is destabilized. If
$\Lambda_{eff} \neq 0 $ it goes to an AdS space while if
$\Lambda_{eff}=  0 $ it goes to flat space.

Finally, we studied the thermodynamics of our hairy charge black
hole solutions. In the case of $\Lambda_{eff}=  0 $ we found that
at all temperature the RN black hole solution is thermodynamically
preferred over the hairy charge black hole solution. In the case
of $\Lambda_{eff} \neq 0 $ we found that the hairy charge black
hole is thermodynamically preferred over the RN black hole at low
temperature. This picture is in agreement with the findings of the
application of the AdS/CFT correspondence to condensed matter
systems. In these systems there is a critical temperature below
which the system undergoes a phase transition to a hairy black
hole configuration at low temperature. This corresponds in the
boundary field theory to the formation of a condensation of the
scalar field.

It would be interesting to study the stability of our solutions.
This stability analysis will make our results robust specially in
connection with the  gauge/gravity applications to condensed
matter systems. Also it would be interesting to extent this study
to a gravitational system with a charged scalar field. In this
case we expect to have a better stability  behaviour of the
system, because the electromagnetic force may balance the
gravitational force. Work in this direction is in progress.

\acknowledgments We would like to thank Theodoros Kolyvaris for
valuable discussions and Cristian Martinez for his valuable
comments and remarks. This work was funded by Comisi\'{o}n
Nacional de Ciencias y Tecnolog\'{i}a through FONDECYT Grants
1110076 (J.S., E.P.) and 11121148 (Y.V.) and by DI-PUCV Grant
123713 (JS). J.S., Y.V. and P.A.G. acknowledge the hospitality of
the National Technical University of Athens where part of this
work was carried out. E.P. is partially supported by ARISTEIA II
action of the operational programme education and long life
learning which is co-funded by the European Union (European Social
Fund) and National Resources. E.P. and J.S. wish to  thank the
Department of Physics, Shanghai Jiao Tong University, and in
particular Prof. Bin Wang for their kind hospitality.

\end{document}